\documentclass[longauth]{aa} 
\usepackage{balance}
\usepackage{wrapfig}
\usepackage{latexsym}                                   
\usepackage{longtable}
\usepackage{graphicx}
\usepackage{natbib}
\usepackage{natbib}
\bibpunct{(}{)}{;}{a}{}{,}
\usepackage{txfonts}
\begin{document}
   \title{The Gaia-ESO Survey: CNO abundances in the open clusters Trumpler\,20, NGC\,4815, and NGC\,6705\thanks{Based on data products from 
   observations made with ESO Telescopes at the La Silla Paranal Observatory under programme ID 188.B-3002 (The Gaia-ESO Public Spectroscopic 
   Survey, PIs G. Gilmore and S. Randich).} }

   \author{G. Tautvai\v{s}ien\.{e}\inst{1},
          A. Drazdauskas\inst{1},
          \v{S}. Mikolaitis\inst{1,2}, 
          G. Barisevi\v{c}ius\inst{1}, 
          E. Puzeras\inst{1}, 
          E. Stonkut\.{e}\inst{1}, 
          Y. Chorniy\inst{1},  
          L. Magrini\inst{3},
          D. Romano\inst{4},
          R. Smiljanic\inst{5}, 
          A. Bragaglia\inst{4}, 
          G. Carraro\inst{6}, 
          E. Friel\inst{7}, 
          T. Morel\inst{8},
          E. Pancino\inst{4,9}, 
          P. Donati\inst{4},
          F. Jim\'enez-Esteban\inst{10},
G. Gilmore\inst{11}, 
S. Randich\inst{3},
R.~D. Jeffries\inst{12},
A. Vallenari\inst{13},
T. Bensby\inst{14},
E. Flaccomio\inst{15},
A. Recio-Blanco\inst{2},
M.~T. Costado\inst{16},
V. Hill\inst{2},
P. Jofr\'e\inst{11},
C. Lardo\inst{17},
P. de Laverny\inst{2},
T. Masseron\inst{11},
L. Moribelli\inst{3},
S. G. Sousa\inst{18}, 
S. Zaggia\inst{4}
}
          
   \institute{Institute of Theoretical Physics and Astronomy, Vilnius University,
              A. Gostauto 12, LT-01108 Vilnius, Lithuania 
              \email{grazina.tautvaisiene@tfai.vu.lt}
   \and Laboratoire Lagrange (UMR7293), Universit\'e de Nice Sophia Antipolis, CNRS,Observatoire de la C\^ote d'Azur, CS 34229, F-06304 Nice cedex 4, France 
   \and INAF - Osservatorio Astrofisico di Arcetri, Largo E. Fermi, 5, 50125 Florence, Italy 
   \and INAF - Osservatorio Astronomico di Bologna, via Ranzani 1, 40127, Bologna, Italy 
   \and Department for Astrophysics, Nicolaus Copernicus Astronomical Center, ul. Rabia\'{n}ska 8, 87-100 Toru\'{n}, Poland 
              \and ESO, Alonso de Cordova 3107, 19001, Santiago de Chile, Chile 
              \and Department of Astronomy, Indiana University, Bloomington, IN 47405, USA 
              \and Institut d'Astrophysique et de G\'eophysique, Universit\'e de Li\`ege, All\'ee du 6 Ao\^ut, B\^at. B5c, 4000 Li\`ege, Belgium 
              \and ASI Science Data Center, Via del Politecnico SNC, 00133 Roma, Italy 
\and Spanish Virtual Observatory, Centro de Astrobiología (INTA-CSIC), P.O. Box 78, 28691 Villanueva de la Cañada, Madrid, Spain 
\and Institute of Astronomy, University of Cambridge, Madingley Road, Cambridge CB3 0HA, United Kingdom
\and Astrophysics Group, Research Institute for the Environment, Physical Sciences and Applied Mathematics, Keele University, Keele, Staffordshire ST5 5BG, United Kingdom
\and INAF - Padova Observatory, Vicolo dell'Osservatorio 5, 35122 Padova, Italy
\and Lund Observatory, Department of Astronomy and Theoretical Physics, Box 43, SE-221 00 Lund, Sweden
\and INAF - Osservatorio Astronomico di Palermo, Piazza del Parlamento 1, 90134, Palermo, Italy
\and Instituto de Astrof\'{i}sica de Andaluc\'{i}a-CSIC, Apdo. 3004, 18080 Granada, Spain
\and Astrophysics Research Institute, Liverpool John Moores University,
146 Brownlow Hill, Liverpool L3 5RF, United Kingdom
\and Centro de Astrof\'isica, Universidade do Porto, Rua das Estrelas, 4150-762 Porto, Portugal Departamento de F\'isica e Astronomia, Faculdade de Ci\^encias, Universidade do Porto, Rua do Campo Alegre, 4169-007 Porto, Portugal
}

   \date{Received September 15, 2014; accepted October 20, 2014}

 
  \abstract
   {The Gaia-ESO  Public Spectroscopic Survey will observe a large sample of clusters and cluster stars, covering a wide 
   age-distance-metallicity-position-density parameter space. }
   {We aim to determine C, N, and O abundances in stars of Galactic open clusters of the Gaia-ESO survey and to compare the observed abundances with those 
predicted by current stellar and Galactic evolution models. In this pilot paper, we investigate the first three intermediate-age open clusters. }
   {High-resolution spectra, observed with the FLAMES-UVES spectrograph on the ESO VLT telescope, were analysed using a differential 
   model atmosphere method. Abundances of carbon were  derived using the ${\rm C}_2$ band heads at
5135 and 5635.5~{\AA}. The wavelength interval 6470--6490~{\AA}, with 
CN features, was analysed to determine nitrogen abundances.  Oxygen abundances were
determined from the [O\,{\sc i}] line at 6300~{\AA}.  }
   { The mean values of the elemental abundances in Trumpler\,20 as determined from 42 stars are: ${\rm [Fe/H]}=0.10\pm0.08$ (s.d.), ${\rm [C/H]}=-0.10\pm0.07$, 
   ${\rm [N/H]}=0.50\pm0.07$, and consequently ${\rm C/N}=0.98\pm0.12$. We measure from five giants in NGC\,4815: ${\rm [Fe/H]}=-0.01\pm0.04$, 
   ${\rm [C/H]}=-0.17\pm0.08$, ${\rm [N/H]}=0.53\pm0.07$, ${\rm [O/H]}=0.12\pm0.09$, and ${\rm C/N}=0.79\pm0.08$. We obtain from 27 giants in NGC\,6705: 
   ${\rm [Fe/H]}=0.0\pm0.05$, ${\rm [C/H]}=-0.08\pm0.06$, ${\rm [N/H]}=0.61\pm0.07$, ${\rm [O/H]}=0.13\pm0.05$, and ${\rm C/N}=0.83\pm0.19$. 
   The C/N ratios of stars in the investigated open clusters were compared with the ratios predicted by stellar evolutionary models.  For the corresponding stellar turn-off masses from 1.9 to 3.3~$M_{\odot}$, the observed C/N ratio values are very close to the predictions of standard first dredge-up models as well as to models of thermohaline extra-mixing.   
   They are not decreased as much as predicted by the recent model in which the thermohaline- and rotation-induced extra-mixing act together. 
   The average [O/H] abundance ratios of NGC\,4815 and NGC\,6705  are compared with the predictions of 
two Galactic chemical evolution models. 
The data are consistent with the evolution at the solar radius within the errors.}
{The first results of CNO determinations in open clusters show the potential of the Gaia-ESO Survey   
to judge stellar and Galactic chemical evolution models and the validity of their physical assumptions through a homogeneous and detailed spectral analysis.   }


\titlerunning{The Gaia-ESO Survey: CNO abundances in the open clusters Trumpler\,20, NGC\,4815, and NGC\,6705}
\authorrunning{G. Tautvai\v{s}ien\.{e} et al. }

\maketitle


\section{Introduction}

Carbon, nitrogen, and oxygen (CNO) are important in Galactic and stellar evolution for many reasons. They 
comprise most of the mass of elements heavier than helium, so their abundances reflect the bulk of chemical enrichment. 
The CNO elements are among the first elements to form in the nucleosynthesis chain. 
These elements play important roles in stellar interiors as sources of opacity and energy production through the CNO cycle, 
and thus affect the star's lifetime, its position in the Hertzsprung-Russell (HR) diagram, and its heavy-element yields. 
The CNO isotopes originate in different stages of the evolution of stars of different masses. In the Galaxy, their relative 
abundances vary spatially and with time. Therefore, they can provide important information about the Galactic chemical  evolution.
Investigating abundances of CNO in objects for which the spatial position and age can be determined with 
the best precision can provide the most valuable information. When studying the evolution of the Galactic disc, such 
objects, without doubt, are open star clusters (c.f. \citealt{Janes79, Panagia80, Cameron85, Friel95, 
Twarog97, Carraro98, Carraro07, Friel02, Chen03, Salaris04,  Bragaglia08, Sestito08, 
Jacobson09, Pancino10, Magrini09, Magrini10, Magrini14, Lepine11, Yong12}, and references therein). 

Open star clusters are even more important in giving us the opportunity to investigate stellar evolution. 
In open clusters we can analyse a number of stars of essentially the same age, distance, and origin, as open cluster stars 
are most likely formed in the same protocloud of gas and dust (see, e.g., \citealt{Lada03, Pallavicini03}). 
There is a wide discussion nowadays to assess if open clusters present singe or multiple populations, however, this 
 question is applicable predominantly to massive open clusters (see \citealt{Cantat14}, and references therein).  
If CNO abundances in cluster members initially were identical, their abundance 
changes in stellar atmospheres of evolved stars  are mainly related to internal processes of stellar evolution. This circumstance 
was exploited in a number of studies of open clusters (\citealt{Gilroy89, Gilroy91,  Luck94, 
Gonzalez00, Tautvaisiene00, Tautvaisiene05,  Origlia06, Smiljanic09, Mikolaitis10, 
Mikolaitis11A, Mikolaitis11B, Mikolaitis12}, among others).
The observational data have provided evidence not only of the first dredge-up (1DUP; \citealt{Iben65, Iben67,  
Dearborn76}), which brings the CN-processed material up to the surfaces of low-mass stars when they reach the bottom of 
the red giant branch (RGB),  but also show evidence of extra-mixing, which happens later on the giant branch.       

  \begin{figure}
   \centering
  \includegraphics[width=0.47\textwidth]{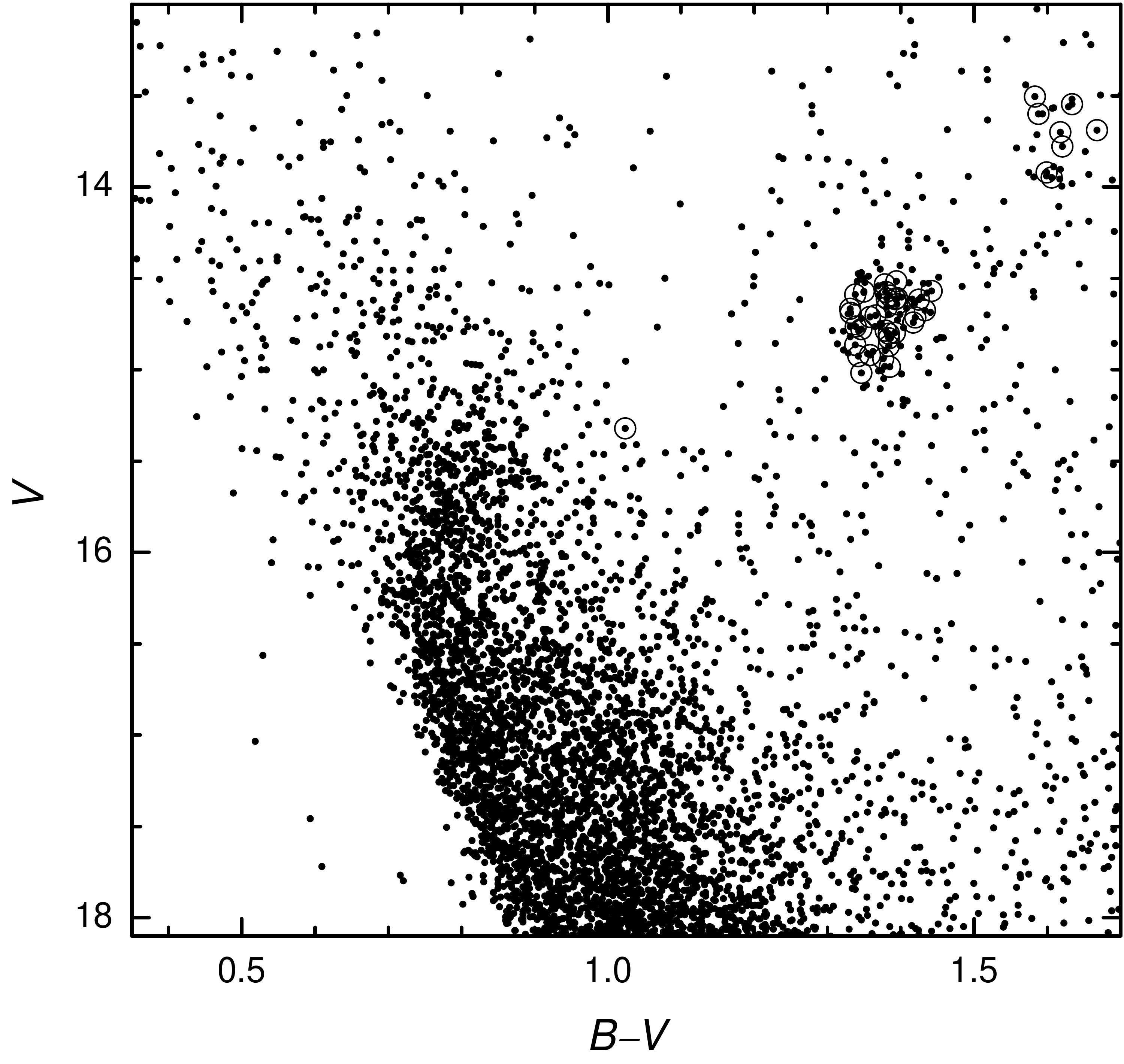}
    \caption{The colour-magnitude diagram of the open cluster Trumpler\,20. The stars investigated in this work are 
indicated by circles. 
The diagram is based on photometry by \citet{Carraro10}.
              }
         \label{Fig.1}
   \end{figure}

  \begin{figure}
   \centering
  \includegraphics[width=0.47\textwidth]{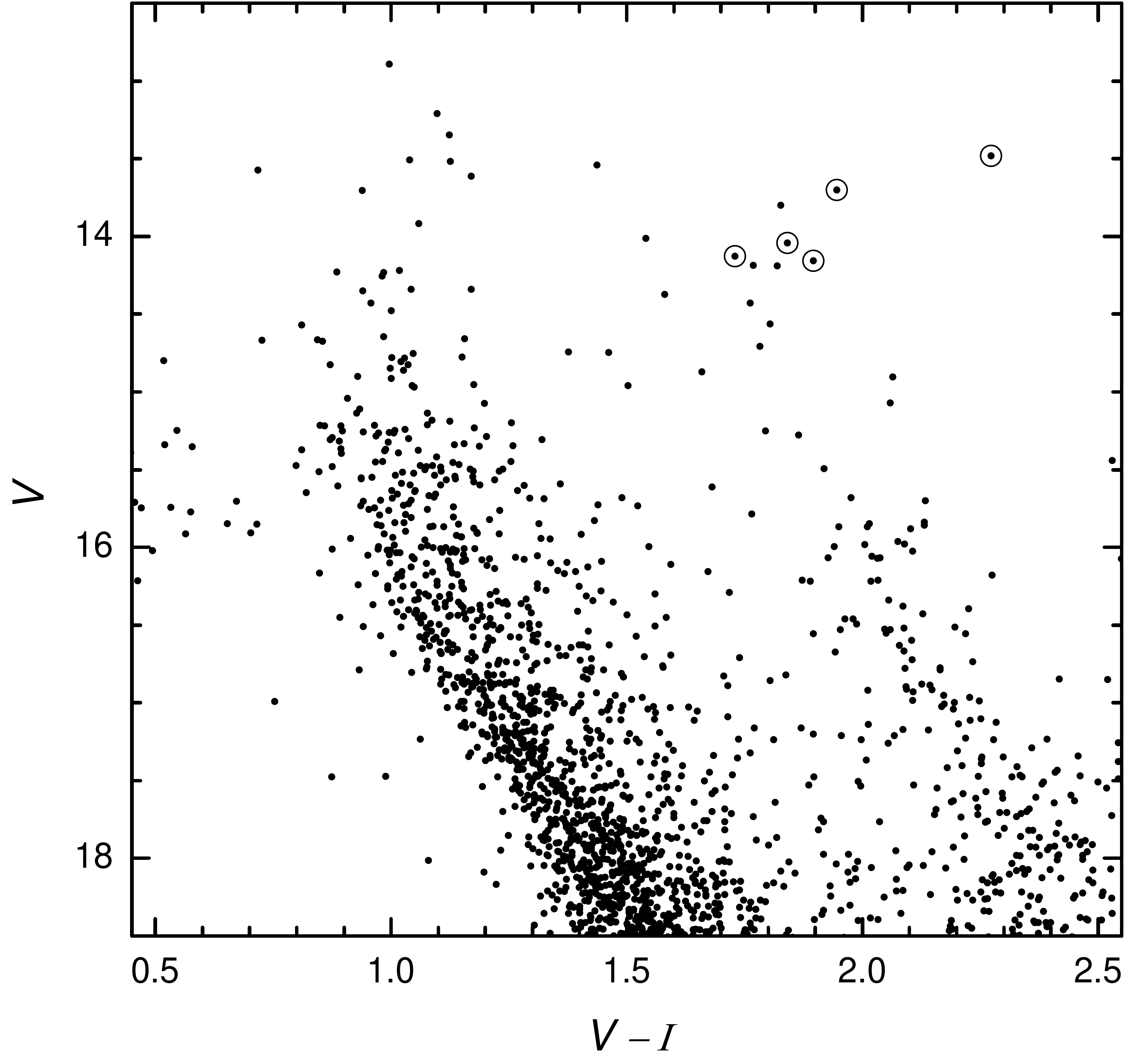}
    \caption{The colour-magnitude diagram of the open cluster NGC\,4815. The stars investigated in this work are 
indicated by circles. 
The diagram is based on photometry by \citet{Prisinzano01}. 
              }
         \label{Fig.2}
   \end{figure}
  \begin{figure}
   \centering
  \includegraphics[width=0.47\textwidth]{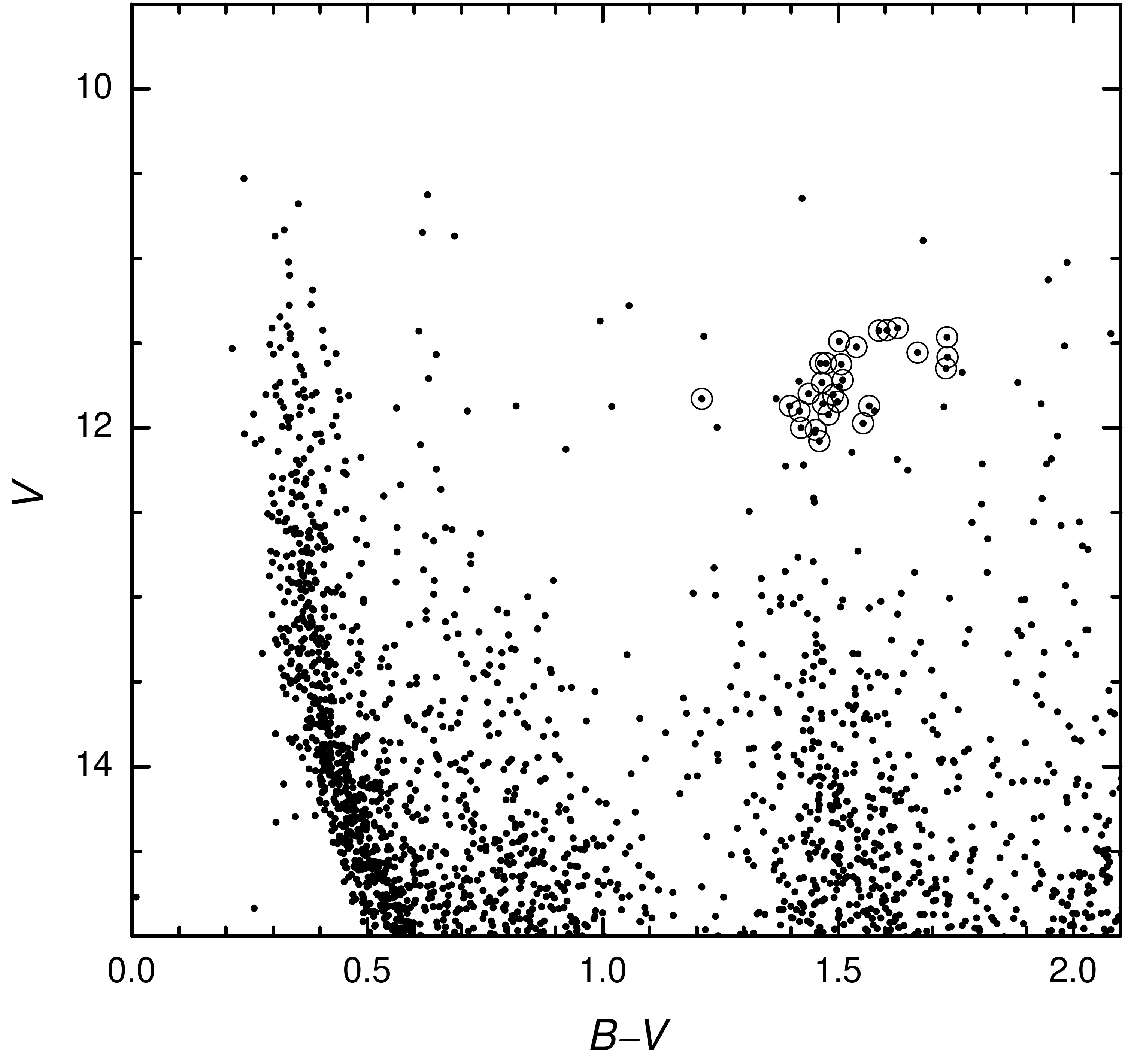}
    \caption{The colour-magnitude diagram of the open cluster NGC\,6705. The stars investigated in this work are 
indicated by circles. 
The diagram is based on photometry by \citet{Sung99}.
              }
         \label{Fig.3}
   \end{figure}

The extra-mixing processes become efficient on the RGB when   
these stars reach the so-called RGB  bump (\citealt{Charbonnel94, Charbonnel98}). It also has been recognised that alterations of surface 
abundances depend on stellar evolutionary stage, mass, and metallicity (e.g. \citealt{Boothroyd99, Gratton00, 
Chaname05,  Charbonnel06, Eggleton06, Cantiello10, Charbonnel10, Tautvaisiene10, Tautvaisiene13, Lagarde12}).  

The nature of the extra mixing itself is still a matter of debate. Currently, thermohaline mixing seems to be the preferred
 mechanism, as it fulfills most of the requirements to explain the observations (see e.g. \citealt{Charbonnel07, Angelou12}). Nevertheless, there are still open questions regarding the physical properties of this mechanism itself (\citealt{Church14}), its 
 efficiency to transport chemicals (see e.g. \citealt{Denissenkov11}), and whether or not it might be suppressed in the presence of other mixing mechanisms (\citealt{Maeder13}). We also note that other extra mixing mechanisms like magnetic buoyancy have been suggested and need verification (\citealt{Busso07, Palmerini11A, Palmerini11B}). 
A better characterisation of these processes needs a comprehensive and statistically significant observational investigation in stars of  different masses 
and metallicities. 

The Gaia-ESO Spectroscopic Survey (GES, see \citealt{Gilmore12, Randich13}) provides an opportunity to address these issues.  The large, 
public spectroscopic survey of the Galaxy using the high-resolution, multi-object spectrograph on the Very Large Telescope 
(ESO, Chile) is targeting about $10^5$ stars in the bulge, thick and thin discs, and halo components, and a sample of up to 80 open clusters 
of various ages, metallicities, locations, and masses.

In this work, we present investigations of CNO abundances in the open clusters Trumpler\,20 (Tr\,20), 
NGC\,4815, and NGC\,6705 (M\,11), which were observed during the first six months of the GES survey. 
All three clusters are located in the inner part of the Galaxy: Trumpler\,20 is at $l = 301.475$ and $b = 2.221$, NGC\,4815 is at $l = 303.6$  
and  $b = -2.1$,  and NGC\,6705 is at $l = 27.307$ and  $b = -2.776$.

Trumpler\,20 is a relatively old open cluster.  
Based on the first internal GES data release (GESviDR1Final) and photometric observations, comparisons with three different sets of isochrones yielded consistent determinations of the cluster age 
of 1.35 to 1.66 Gyr, a turn-off mass of $1.9 \pm 0.1 M_{\odot}$, a distance of $3.4 - 3.5$~kpc, the Galactocentric radius ($R_{\rm GC}$) of 6.88~kpc, and the average iron abundance of 13 members was ${\rm [Fe/H]}=0.17$ (\citealt{Donati14}). 
  
NGC\,4815 is an intermediate age cluster. It  has not been observed spectroscopically before GES. 
According to GESviDR1Final, a mean [Fe/H] was determined to be equal to $+0.03 \pm 0.05$~ dex. Comparisons with three different sets of isochrones 
yielded consistent determinations of the cluster age of 0.5 to 0.63 Gyr, a turn-off mass of $2.6 \pm 0.1M_{\odot}$, a distance of $2.5 - 2.7$~kpc,  
and  $R_{\rm GC}=6.9$~kpc (\citealt{Friel14}). 

NGC\,6705 is the youngest among the open clusters in this paper. As determined by \citet{Cantat14} on the basis of GESviDR1Final,  the age of NGC\,6705 is in 
the range from 0.25 to 0.32~Gyr, the turn-off mass from 3.47 to 3.2~$M_{\odot}$, depending on the adopted stellar model, and the Galactocentric radius equal 
to 6.3~kpc. The average iron abundance of 21 members was ${\rm [Fe/H]}=0.10\pm0.06$.  

This pilot paper on CNO abundances in the first three intermediate-age open clusters is based on the second internal GES data release (GESviDR2Final).

 \begin{table*}
     \centering
\begin{minipage}{150mm}
\caption{Stellar parameters for target stars in Trumpler\,20}
\label{table:1}
\begin{tabular}{rccccccrcc}
\hline\hline
\noalign{\smallskip}
ID &	GES ID &	R.A. & DEC & \emph{V} & \emph{B -- V} & RV & S/N & \emph{v} sin \emph{i}~\tablefootmark{a}  & \emph{v} sin \emph{i}~\tablefootmark{b}  \\
	&	&	deg (J2000)	&	deg (J2000)	&	mag	&	mag	&	km s$^{-1}$ 	&		&	km s$^{-1}$ 	&	km s$^{-1}$ 	\\
\hline																			
\noalign{\smallskip}																			
2730	&	12383595-6045245	&	189.6498	&	--60.7568	&	14.942	&	1.376	&	--41.39	&	33	&	2.00	&	3.0	\\
2690	&	12383659-6045300	&	189.6525	&	--60.7583	&	14.572	&	1.442	&	--42.57	&	20	&	2.46	&	3.5	\\
63	&	12385807-6030286	&	189.7420	&	--60.5079	&	13.603	&	1.588	&	--42.35	&	71	&	2.61	&	3.5	\\
292	&	12390411-6034001	&	189.7671	&	--60.5667	&	13.506	&	1.583	&	--40.45	&	72	&	2.64	&	2.6	\\
1082	&	12390478-6041475	&	189.7699	&	--60.6965	&	14.800	&	1.392	&	--41.78	&	36	&	2.00	&	3.0	\\
724	&	12390710-6038057	&	189.7796	&	--60.6349	&	15.018	&	1.346	&	--41.44	&	31	&	2.00	&	2.5	\\
794	&	12391004-6038402	&	189.7918	&	--60.6445	&	13.703	&	1.618	&	--41.89	&	57	&	3.30	&	4.0	\\
582	&	12391114-6036527	&	189.7964	&	--60.6146	&	14.923	&	1.358	&	--42.66	&	36	&	2.00	&	3.0	\\
542	&	12391201-6036322	&	189.8000	&	--60.6089	&	14.690	&	1.332	&	--41.80	&	40	&	2.00	&	2.5	\\
340	&	12391577-6034406	&	189.8157	&	--60.5779	&	14.669	&	1.331	&	--40.01	&	36	&	2.00	&	3.5	\\
770	&	12392585-6038279	&	189.8577	&	--60.6411	&	14.928	&	1.342	&	--42.60	&	23	&	2.00	&	3.5	\\
950	&	12392637-6040217	&	189.8599	&	--60.6727	&	14.831	&	1.384	&	--41.85	&	27	&	2.36	&	3.5	\\
505	&	12392700-6036053	&	189.8625	&	--60.6015	&	14.519	&	1.394	&	--40.02	&	46	&	2.00	&	3.5	\\
894	&	12393132-6039422	&	189.8805	&	--60.6617	&	14.766	&	1.339	&	--36.09	&	40	&	2.25	&	3.5	\\
203	&	12393741-6032568	&	189.9059	&	--60.5491	&	14.865	&	1.338	&	--41.44	&	29	&	2.00	&	3.0	\\
835	&	12393782-6039051	&	189.9076	&	--60.6514	&	14.577	&	1.380	&	--40.69	&	32	&	2.00	&	4.5	\\
1010	&	12394051-6041006	&	189.9188	&	--60.6835	&	14.643	&	1.395	&	--42.79	&	32	&	2.26	&	3.5	\\
923	&	12394122-6040040	&	189.9218	&	--60.6678	&	14.872	&	1.384	&	--39.32	&	37	&	2.00	&	3.5	\\
858	&	12394309-6039193	&	189.9295	&	--60.6554	&	14.676	&	1.433	&	--41.82	&	31	&	2.00	&	3.5	\\
227	&	12394387-6033166	&	189.9328	&	--60.5546	&	14.590	&	1.338	&	--41.00	&	33	&	2.00	&	5.0	\\
346	&	12394419-6034412	&	189.9341	&	--60.5781	&	14.704	&	1.366	&	--41.08	&	35	&	2.00	&	3.5	\\
781	&	12394475-6038339	&	189.9365	&	--60.6428	&	14.610	&	1.395	&	--39.37	&	36	&	2.00	&	3.5	\\
768	&	12394517-6038257	&	189.9382	&	--60.6405	&	14.789	&	1.380	&	--41.50	&	38	&	2.00	&	3.5	\\
791	&	12394596-6038389	&	189.9415	&	--60.6441	&	14.535	&	1.378	&	--39.72	&	44	&	2.00	&	3.0	\\
287	&	12394690-6033540	&	189.9454	&	--60.5650	&	14.781	&	1.346	&	--40.99	&	37	&	2.00	&	3.5	\\
1008	&	12394715-6040584	&	189.9465	&	--60.6829	&	13.949	&	1.606	&	--40.30	&	54	&	2.87	&	3.5	\\
795	&	12394742-6038411	&	189.9476	&	--60.6448	&	14.712	&	1.358	&	--40.00	&	32	&	2.00	&	4.0	\\
246	&	12394899-6033282	&	189.9541	&	--60.5578	&	14.575	&	1.350	&	--39.67	&	30	&	2.33	&	4.0	\\
787	&	12395426-6038369	&	189.9761	&	--60.6436	&	14.597	&	1.385	&	--42.45	&	45	&	2.00	&	4.0	\\
638	&	12395555-6037268	&	189.9815	&	--60.6241	&	14.613	&	1.381	&	--41.06	&	57	&	2.00	&	3.0	\\
430	&	12395569-6035233	&	189.9820	&	--60.5898	&	15.322	&	1.024	&	--38.99	&	25	&	4.25	&	8.0	\\
827	&	12395655-6039011	&	189.9856	&	--60.6503	&	14.805	&	1.384	&	--38.99	&	39	&	2.00	&	3.0	\\
885	&	12395712-6039335	&	189.9880	&	--60.6593	&	14.660	&	1.381	&	--42.06	&	42	&	2.00	&	3.0	\\
399	&	12395975-6035072	&	189.9990	&	--60.5853	&	14.618	&	1.425	&	--42.10	&	53	&	2.00	&	4.0	\\
129	&	12400111-6031395	&	190.0046	&	--60.5276	&	14.716	&	1.420	&	--40.90	&	39	&	2.00	&	3.0	\\
429	&	12400116-6035218	&	190.0048	&	--60.5894	&	14.615	&	1.393	&	--40.87	&	43	&	2.00	&	4.0	\\
911	&	12400260-6039545	&	190.0108	&	--60.6651	&	13.780	&	1.621	&	--41.29	&	72	&	2.82	&	4.0	\\
1044	&	12400278-6041192	&	190.0116	&	--60.6887	&	14.985	&	1.385	&	--39.09	&	27	&	2.00	&	3.0	\\
591	&	12400451-6036566	&	190.0188	&	--60.6157	&	13.690	&	1.668	&	--41.65	&	62	&	3.49	&	4.0	\\
468	&	12400755-6035445	&	190.0315	&	--60.5957	&	13.548	&	1.634	&	--41.12	&	60	&	3.51	&	4.5	\\
679	&	12402229-6037419	&	190.0929	&	--60.6283	&	14.743	&	1.417	&	--41.55	&	31	&	2.30	&	5.0	\\
3470	&	12402480-6043101	&	190.1033	&	--60.7195	&	13.922	&	1.599	&	--40.70	&	47	&	3.08	&	4.0	\\
\hline																			
																			
\end{tabular}												
\end{minipage}
\tablefoot{ID and photometric data were taken from \citet{Carraro10}, \tablefoottext{a}{values from the GES database}, 
\tablefoottext{b}{determined in this work.}} 
\end{table*} 


 \begin{table*}
     \centering
\begin{minipage}{150mm}
\caption{Stellar parameters for target stars in NGC\,4815}
\label{table:2}
\begin{tabular}{rccccccrcc}
\hline\hline
\noalign{\smallskip}
ID &	GES ID &	R.A. & DEC & \emph{V} & \emph{V -- I} & RV & S/N	& \emph{v} sin \emph{i}~\tablefootmark{a} 	& \emph{v} sin \emph{i}~\tablefootmark{b} 	\\
	&	&	deg (J2000)	&	deg (J2000)	&	mag	&	mag	&	km s$^{-1}$ 	&		&	km s$^{-1}$ 	&	km s$^{-1}$ 	\\
\hline																			
\noalign{\smallskip}																			
1795	&	12572442-6455173	&	194.3518	&	--64.9215	&	13.482	&	2.273	&	--30.11	&	65	&	4.56	&	5.0	\\
358	&	12574328-6457386	&	194.4303	&	--64.9607	&	13.703	&	1.946	&	--31.12	&	63	&	4.09	&	5.5	\\
210	&	12575511-6458483	&	194.4796	&	--64.9801	&	14.043	&	1.841	&	--29.92	&	53	&	3.25	&	4.5	\\
95	&	12575529-6456536	&	194.4804	&	--64.9482	&	14.128	&	1.730	&	--30.75	&	49	&	2.00	&	3.5	\\
106	&	12580262-6456492	&	194.5109	&	--64.9470	&	14.158	&	1.896	&	--30.60	&	46	&	3.30	&	6.0	\\
\hline																			
																			
\end{tabular}												
\end{minipage}
\tablefoot{$V$ and $V-I$ were taken from \citet{Prisinzano01}, \tablefoottext{a}{values from the GES database}, 
\tablefoottext{b}{determined in this work.}} 
\end{table*} 

 \begin{table*}
     \centering
\begin{minipage}{150mm}
\caption{Stellar parameters for target stars in NGC\,6705}
\label{table:3}
\begin{tabular}{rccccccrcr}
\hline\hline
\noalign{\smallskip}
ID & GES ID &	R.A.	& DEC &	\emph{V}	& \emph{B -- V}	& RV & S/N & \emph{v}\,sin\,\emph{i}~\tablefootmark{a}  & \emph{v}\,sin\,\emph{i}~\tablefootmark{b} \\
	&	&	deg (J2000)	&	deg (J2000)	&	mag	&	mag	&	km s$^{-1}$ 	&		&	km s$^{-1}$ 	&	km s$^{-1}$ 	\\
\hline																			
\noalign{\smallskip}																			
2000 	&	18502831-0615122	&	282.6180	&	--6.2534	&	11.428	&	1.587	&	34.42	&	81	&	4.02	&	4.5	\\
1837 	&	18503724-0614364	&	282.6552	&	--6.2434	&	12.001	&	1.422	&	35.28	&	131	&	7.89	&	10.5	\\
1658 	&	18504563-0612038	&	282.6901	&	--6.2011	&	11.622	&	1.475	&	35.13	&	111	&	5.40	&	7.0	\\
1625 	&	18504737-0617184	&	282.6974	&	--6.2884	&	11.652	&	1.729	&	32.01	&	102	&	3.84	&	4.5	\\
1446 	&	18505494-0616182	&	282.7289	&	--6.2717	&	11.860	&	1.468	&	34.76	&	104	&	5.24	&	7.0	\\
1423 	&	18505581-0618148	&	282.7325	&	--6.3041	&	11.414	&	1.627	&	35.23	&	121	&	4.73	&	6.0	\\
1364 	&	18505755-0613461	&	282.7398	&	--6.2295	&	11.830	&	1.211	&	30.60	&	162	&	9.14	&	11.5	\\
1286 	&	18505944-0612435	&	282.7477	&	--6.2121	&	11.872	&	1.398	&	34.38	&	117	&	8.25	&	11.0	\\
1256 	&	18510023-0616594	&	282.7510	&	--6.2832	&	11.586	&	1.733	&	35.23	&	88	&	3.54	&	5.0	\\
1248 	&	18510032-0617183	&	282.7513	&	--6.2884	&	12.081	&	1.461	&	35.23	&	122	&	6.48	&	9.0	\\
1184 	&	18510200-0617265	&	282.7583	&	--6.2907	&	11.426	&	1.604	&	32.03	&	81	&	3.68	&	4.5	\\
1145 	&	18510289-0615301	&	282.7620	&	--6.2584	&	12.014	&	1.453	&	32.76	&	118	&	5.43	&	7.0	\\
1117 	&	18510341-0616202	&	282.7642	&	--6.2723	&	11.801	&	1.438	&	36.14	&	105	&	8.03	&	10.0	\\
1111 	&	18510358-0616112	&	282.7649	&	--6.2698	&	11.902	&	1.418	&	34.63	&	113	&	5.31	&	7.5	\\
1090 	&	18510399-0620414	&	282.7666	&	--6.3448	&	11.872	&	1.566	&	34.01	&	102	&	4.65	&	6.0	\\
963 	&	18510662-0612442	&	282.7776	&	--6.2123	&	11.720	&	1.510	&	33.23	&	113	&	5.04	&	7.0	\\
916	&	18510786-0617119	&	282.7828	&	--6.2866	&	11.621	&	1.463	&	33.86	&	188	&	7.62	&	9.5	\\
899	&	18510833-0616532	&	282.7847	&	--6.2814	&	11.736	&	1.466	&	33.14	&	94	&	4.64	&	5.5	\\
827	&	18511013-0615486	&	282.7922	&	--6.2635	&	11.493	&	1.503	&	36.88	&	115	&	3.51	&	4.5	\\
816	&	18511048-0615470	&	282.7937	&	--6.2631	&	11.627	&	1.507	&	33.08	&	121	&	7.45	&	9.0	\\
779	&	18511116-0614340	&	282.7965	&	--6.2428	&	11.468	&	1.732	&	33.42	&	87	&	4.61	&	5.5	\\
686	&	18511452-0616551	&	282.8105	&	--6.2820	&	11.923	&	1.480	&	35.15	&	97	&	7.27	&	9.0	\\
669	&	18511534-0618359	&	282.8139	&	--6.3100	&	11.974	&	1.553	&	33.75	&	161	&	7.13	&	9.0	\\
660	&	18511571-0618146	&	282.8155	&	--6.3041	&	11.807	&	1.490	&	35.17	&	139	&	4.58	&	6.0	\\
411	&	18512662-0614537	&	282.8609	&	--6.2483	&	11.559	&	1.669	&	33.90	&	128	&	5.41	&	6.5	\\
160	&	18514034-0617128	&	282.9181	&	--6.2869	&	11.525	&	1.539	&	33.25	&	121	&	5.26	&	7.0	\\
136	&	18514130-0620125	&	282.9221	&	--6.3368	&	11.849	&	1.499	&	33.21	&	83	&	2.84	&	4.0	\\
\hline																			
\end{tabular}												
\end{minipage}
\tablefoot{ID and photometric data were taken from \citet{Sung99}, \tablefoottext{a}{values from the GES database}, 
\tablefoottext{b}{determined in this work.}} 
\end{table*}

\section{Observations and method of analysis}

\subsection{Observations}

Observations were conducted with the FLAMES (Fiber Large Array Multi-Element Spectrograph) multi-fiber facility (\citealt{Pasquini02}) in spring 
of 2012 and 2013. Spectra of high-resolving power  ($R\approx$ 47\,000) were obtained with UVES (Ultraviolet and Visual Echelle Spectrograph, 
\citealt{Dekker00}).
The spectra were exposed onto two CCDs, resulting in a wavelength coverage of 4700--6840~{\AA} with a gap of about 50~{\AA}  in the centre. 
The spectra were reduced with the ESO UVES pipeline and dedicated scripts described by \citet{Sacco14}. Radial velocities (RV) 
and rotation velocities ($v\,{\rm sin}\,i$) were also determined by cross-correlating all the spectra with a sample of synthetic templates specifically 
derived for the Gaia-ESO project. 

The information on radial velocities was particularly useful in determining true members of the stellar clusters. The typical error on RVs is about 
0.4~km\,s$^{-1}$. In this work, we investigate cluster stars identified by \citet{Donati14} for Trumpler\,20,  by \citet{Friel14} for NGC\,4815, 
and by \citet{Cantat14} for NGC\,6705.  In cases where additional cluster stars in the Gaia-ESO Survey were observed later, the identical 
 mean radial velocity and dispersion values determined in these studies ($-40.26 \pm 0.11$~km\,s$^{-1}$ for Trumpler\,20, and 
$34.1 \pm 1.5$~km\,s$^{-1}$ for NGC\,6705) as well as information of proper motions guided us in identifying the cluster members.  A list of the investigated 42 stars in Trumpler\,20,    
5 stars in NGC\,4815, and  27 stars in NGC\,6705 as well as some of their parameters  are presented in Tables~1--3, respectively. 
Figs.~1--3 show the investigated stars in the colour-magnitude diagrams.   

   \begin{figure*}
   \centering
   \includegraphics[width=0.7\textwidth]{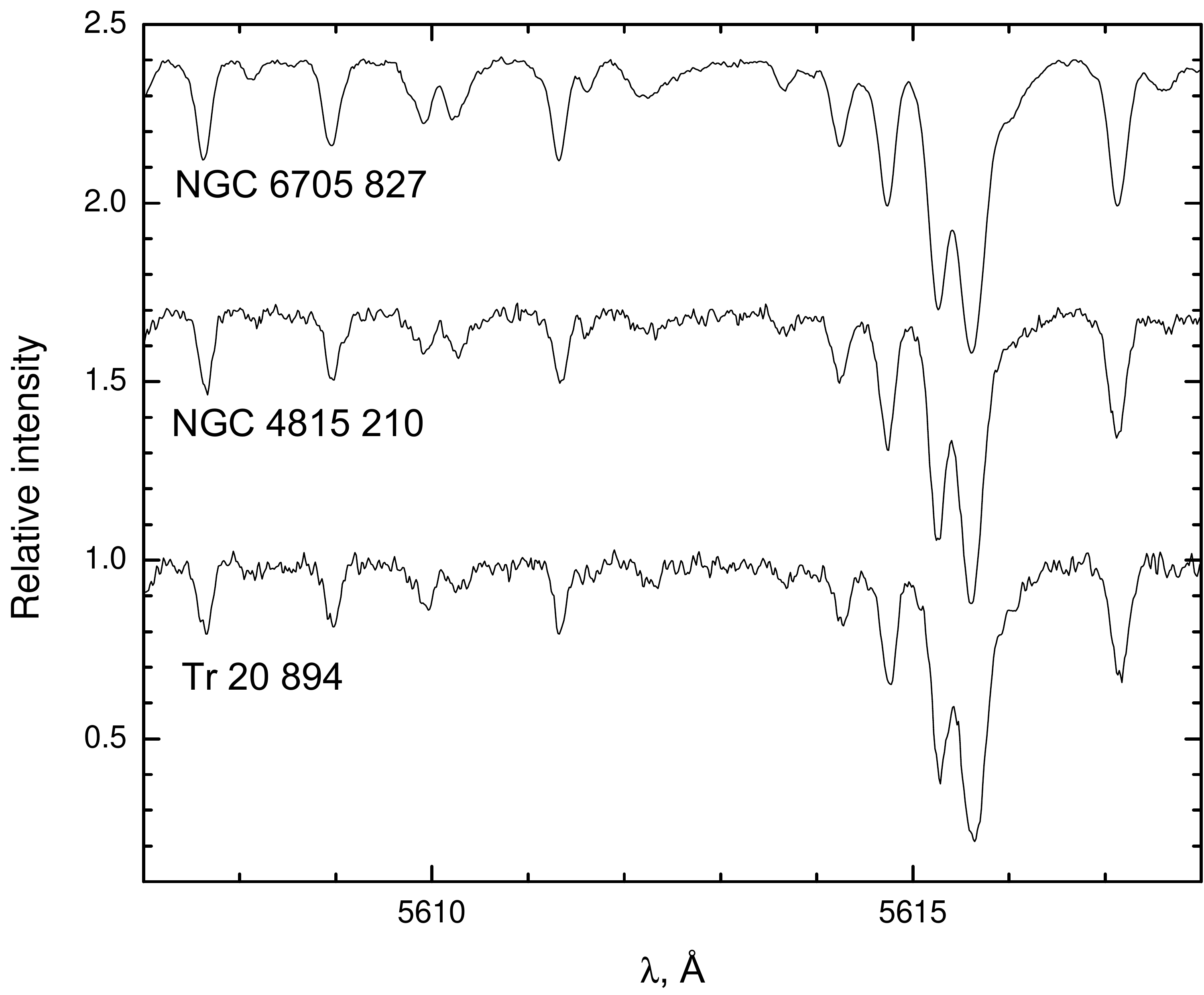}
      \caption{Examples of stellar spectra for our programme stars. An offset of 0.7 in relative flux is applied for clarity.
              }
         \label{Fig.4}
   \end{figure*}

The signal-to-noise ratio (S/N) in the spectra of the observed cluster stars varies depending on their brightness.  The highest S/N, ranging from 80 to 190, 
was achieved for stars in NGC\,6705, from 45 to 65 in NGC\,4815, and from 20 to 70 in Trumpler\,20.  Examples of  stellar spectra with typical S/N for 
stars in each of the clusters are presented in Fig.~4.  All the spectra were taken from the GES 
database\footnote{The operational database has been developed by the Cambridge Astronomical Survey Unit (CASU) based at the Institute of Astronomy at 
the University of Cambridge. See the website http://casu.ast.cam.ac.uk/gaiaeso/ for more information.}. Additional efforts were applied in improving their 
continuum normalisation. For this purpose we used the SPLAT-VO code\footnote{http://star-www.dur.ac.uk/~pdraper/splat/splat-vo/splat-vo.html}.  

In the standard procedure of the $1^{st}$ and $2^{nd}$ GES data releases, only a single, final
file for each star was used for analysis. This file is a sum of all
spectra of a particular star taken in all observing runs (see \citealt{Sacco14}, for more details).
A correction for telluric contamination is not yet implemented, so none is
applied before co-addition. This may cause a problem in some cases,
as we encountered for the [O I] 6300.3~\AA line in stars of Trumpler\,20, because of the systemic radial
velocity of that cluster.  
The oxygen line in the spectra of NGC\,4815 and NGC\,6705 stars was not affected by telluric lines.

\subsection{Main atmospheric parameters}

The main atmospheric parameters of the stars were determined spectroscopically using a technique described by  
\citet{Smiljanic14}.  To make full use of the available expertise of the consortium, all the spectra were analysed in parallel by 13 nodes 
of scientists. The methodology and codes used by each Node are described in detail by \citet{Smiljanic14}.  A number of constraints 
have been imposed on input data used in the analysis to guarantee some degree of homogeneity in the final results: the use of a common 
line list, the use of one single set of model atmospheres, and the analysis of common calibration targets. The Gaia-ESO Survey line list version 4.0 was 
used to determine the main atmospheric parameters of the cluster stars (Heiter et al. 2014, in prep.). 
For model atmospheres, the MARCS grid (\citealt{Gustafsson08}) was adopted. The grid consists of spherically-symmetric models complemented by 
plane-parallel models for stars of high surface gravity (between log g = 3.0 and 5.0, or 5.5 for cooler models). The models  are based on hydrostatic equilibrium, LTE, 
and the mixing-length theory of convection. The assumed solar abundances are those of \citet{Grevesse07}. Thus, when metallicities in the 
format [El/H] or [El/Fe]\footnote{We use the customary spectroscopic notation
[X/Y]$\equiv \log_{10}(N_{\rm X}/N_{\rm Y})_{\rm star} -\log_{10}(N_{\rm X}/N_{\rm Y})_\odot$.}  are quoted in this work, the solar elemental abundances 
are the following: $A{\rm (Fe)}_{\odot}=7.45$, $A{\rm (C)}_{\odot}=8.39$, $A{\rm (N)}_{\odot}=7.78$, and $A{\rm (O)}_{\odot}=8.66$. 

To homogenise the results of different nodes and quantify the method-to-method dispersion of each parameter the median and the associated median 
absolute deviation (MAD) have been used. The first step was a zeroth-order quality control of the results of each node (values with very large error bars 
were removed). Then, we used results of the benchmark stars to weight the performance of each node in different regions of the parameter space.  
Finally, the weighted-median value of the validated results was adopted as the recommended value of that parameter. 
For the Internal Data Release~2 (GESviDR2Final) results, used in this work, the median of the method-to-method dispersion is 55~K, 0.13~dex, and 0.07~dex 
for $T_{\rm eff}$, log~$g$, and [Fe/H], respectively. 

\subsection{C, N, and O abundances}

Abundances of carbon, nitrogen, and oxygen were determined by the GES Vilnius node using the  spectral 
synthesis with the code BSYN, developed at the Uppsala Astronomical Observatory. 
The ${\rm C}_2$ Swan (1,0) band head at 5135~{\AA} and ${\rm C}_2$ Swan (0,1) band head at 5635.5~{\AA} were 
investigated in order to determine 
the carbon abundance.  
The ${\rm C}_2$ bands are suitable for carbon abundance investigations since yield give the same 
carbon abundances as [C\,{\sc i}] 
lines, which are not sensitive to non-local thermodynamical equilibrium (NLTE) deviations (c.f. \citealt{Clegg81, Gustafsson99}).

The interval 6470 -- 6490~{\AA} containing $^{12}{\rm C}^{14}{\rm N}$ bands 
was used for the nitrogen abundance analysis. 
The oxygen abundance was determined from the forbidden [O\,{\sc i}] line at 6300.31~\AA. 
Following \citet{Johansson03}, we took into account the oscillator strength values for \textsuperscript{58}Ni and 
 \textsuperscript{60}Ni, which blend the oxygen line. 
Lines of [O\,{\sc i}] are considered as very good indicators of oxygen abundances. It was 
determined that they are 
not only insensitive to NLTE effects, but also give similar oxygen abundance results with 3D and 1D model 
atmospheres (c.f. \citealt{Asplund04, Pereira09}).   

All the synthetic spectra have been calibrated to the solar spectrum by \citet{Kurucz05} to make the analysis strictly differential. 
Figs.~ 5--8 display examples of spectrum syntheses for the programme stars. The best-fit abundances were 
determined by eye.

In fitting the observed spectra with theoretical spectra the stellar rotation was taken into account.
Approximate values of stellar rotation velocities were evaluated for the Survey stars as described by \citet{Sacco14}.  Values of 
$v\,{\rm sin}\,i$ were calculated using an empirical relation of a full width half maximum of the the cross- correlation function (CCF$_{\rm FWHM}$) 
and $v\,{\rm sin}\,i$, which was specifically derived for this project. Accuracy of those values depends on the spectral type of 
a star. Since the values of $v\,{\rm sin}\,i$, provided by the Gaia-ESO Survey in GESviDR2Final, were estimated before the main atmospheric parameters of stars 
were known, consequently $v\,{\rm sin}\,i$ can be improved after the stellar parameters are determined.  Thus we did that for the investigated stars using stronger surrounding lines in 
spectral regions around the investigated C, N, and O features. The initial and updated $v\,{\rm sin}\,i$ values are presented in the last two columns of Tables~1 to 3. The new values on average are larger by about $1.4\pm0.7$~km\,s$^{-1}$. 
 We did not need an extremely high accuracy of the $v\,{\rm sin}\,i$ values since their influence in determining of the C, N, and O abundances is not crucial 
(see Sect.~2.4). However, especially for stars of our youngest open cluster NGC\,6705, rotating up to 11~km\,s$^{-1}$, a higher accuracy of elemental abundances was certainly achieved. 
    
\begin{figure}
\centering
\includegraphics[height = 6cm]{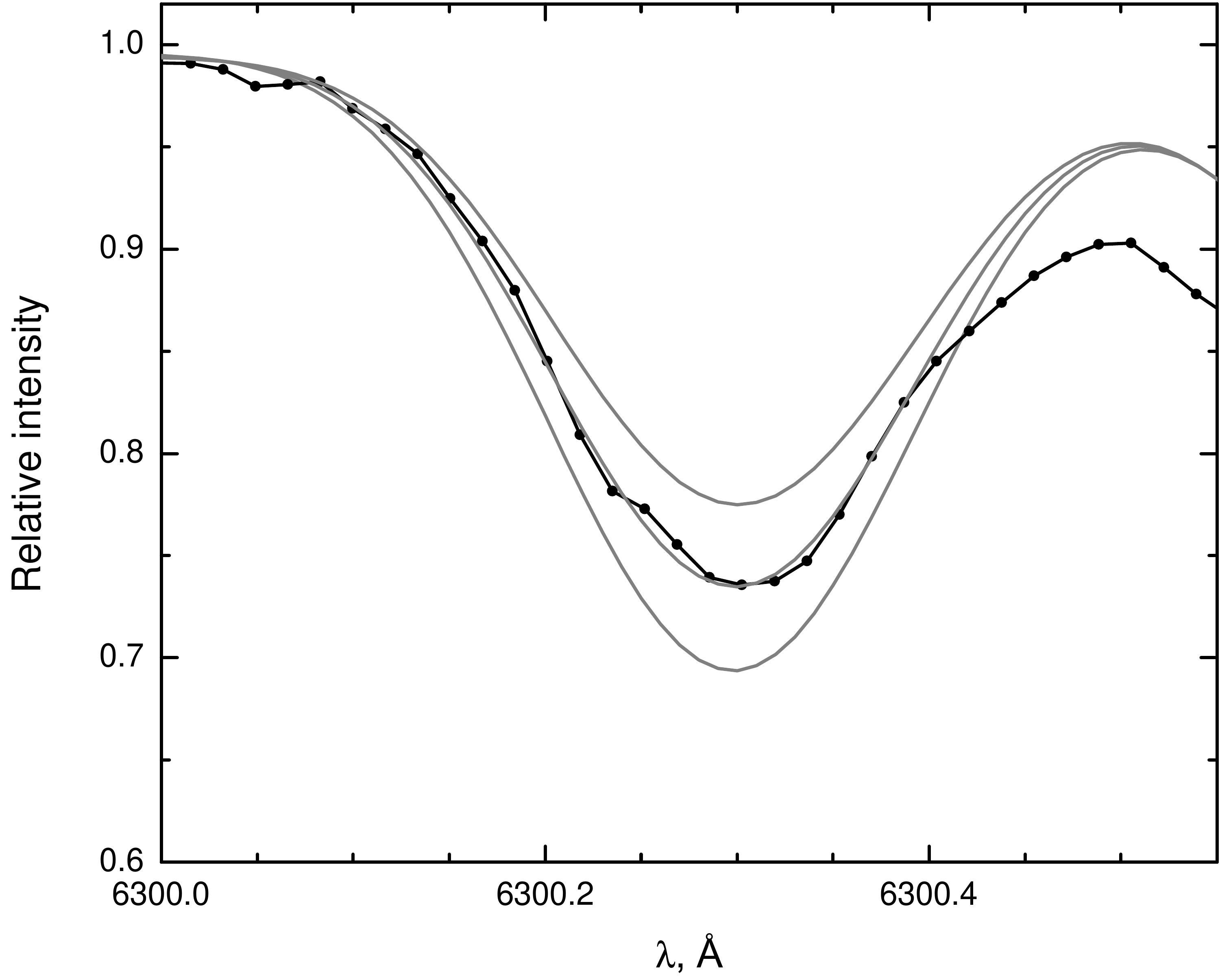}

\caption{A fit to the forbidden [O\,{\sc i}] line at 6300.3 {\AA} in the programme star NGC\,6705\,779. 
The observed spectrum is shown as a black line with dots. The synthetic spectra with ${\rm [O/Fe]}=0.07 \pm 0.1$ 
are shown as grey lines.}
    \label{Fig5}
\end{figure}
%

\begin{figure*}
\centering
\includegraphics[height = 6.5cm]{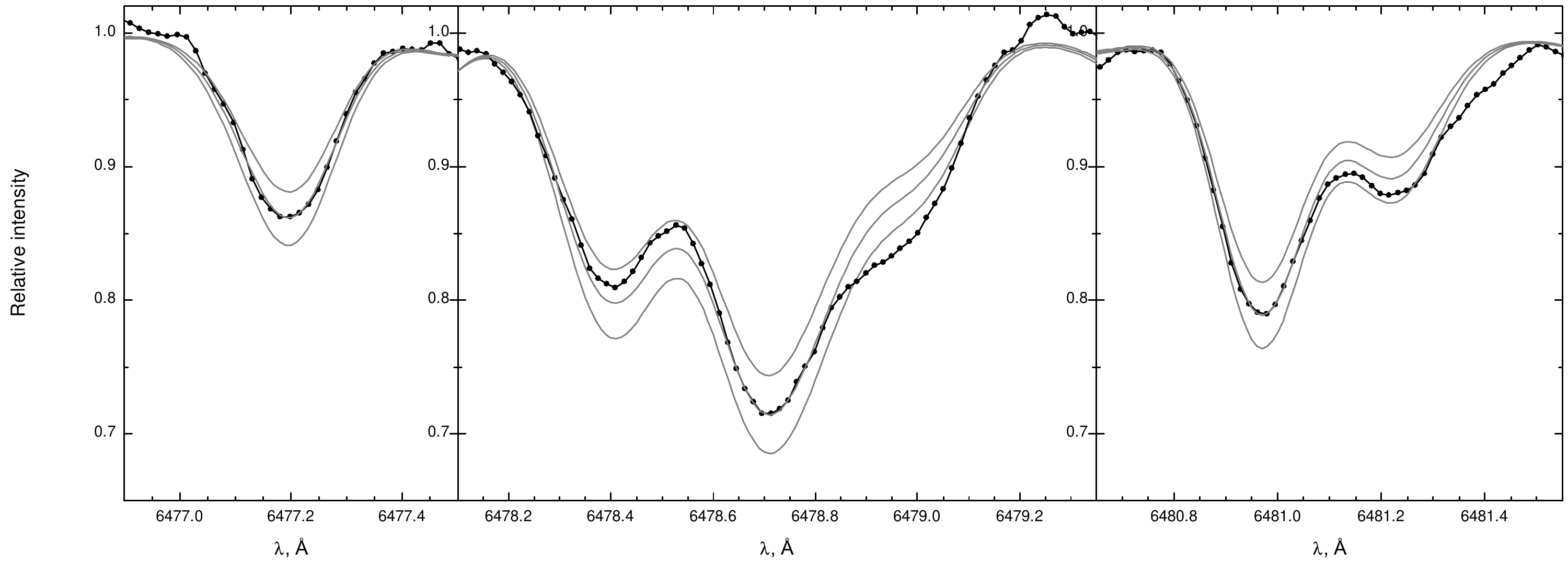}
\caption{A fit to the CN bands in the programme star NGC\,6705\,1625. 
The observed spectrum is shown as a black line with dots. The synthetic spectra with ${\rm [N/Fe]}=0.66 \pm 0.1$ 
are shown as grey lines.}
    \label{Fig6}
\end{figure*}
%

\begin{figure}
\centering
\includegraphics[height = 6cm]{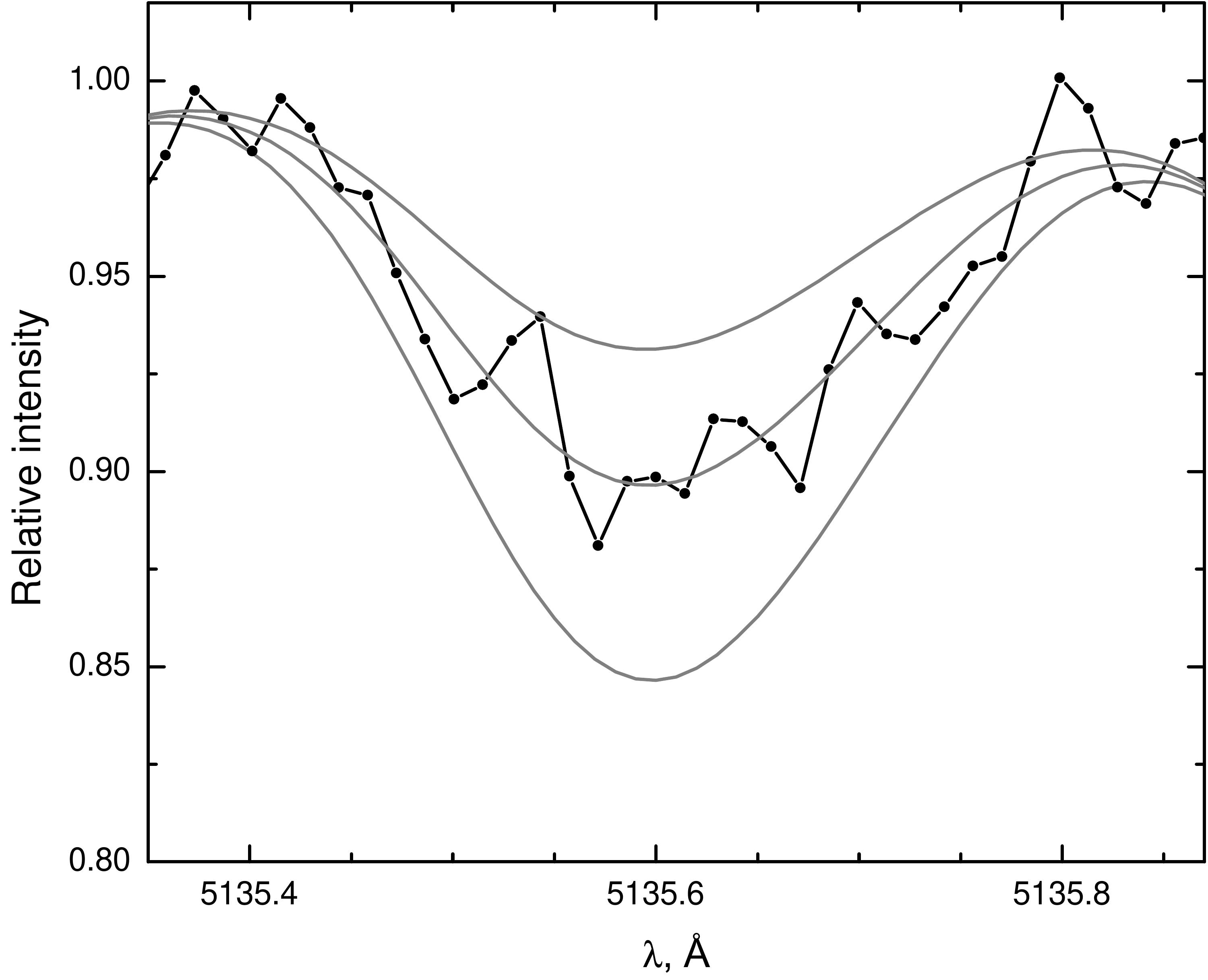}
\caption{A fit to the ${\rm C}_2$ Swan (1,0) band head at 5135~{\AA} in the programme star Tr\,20\,63. 
The observed spectrum is shown as a black line with dots. The synthetic spectra with ${\rm [C/Fe]}=-0.22 \pm 0.1$ 
are shown as grey lines.}
    \label{Fig7}
\end{figure}
%

\begin{figure}
\centering
\includegraphics[height = 6.8cm]{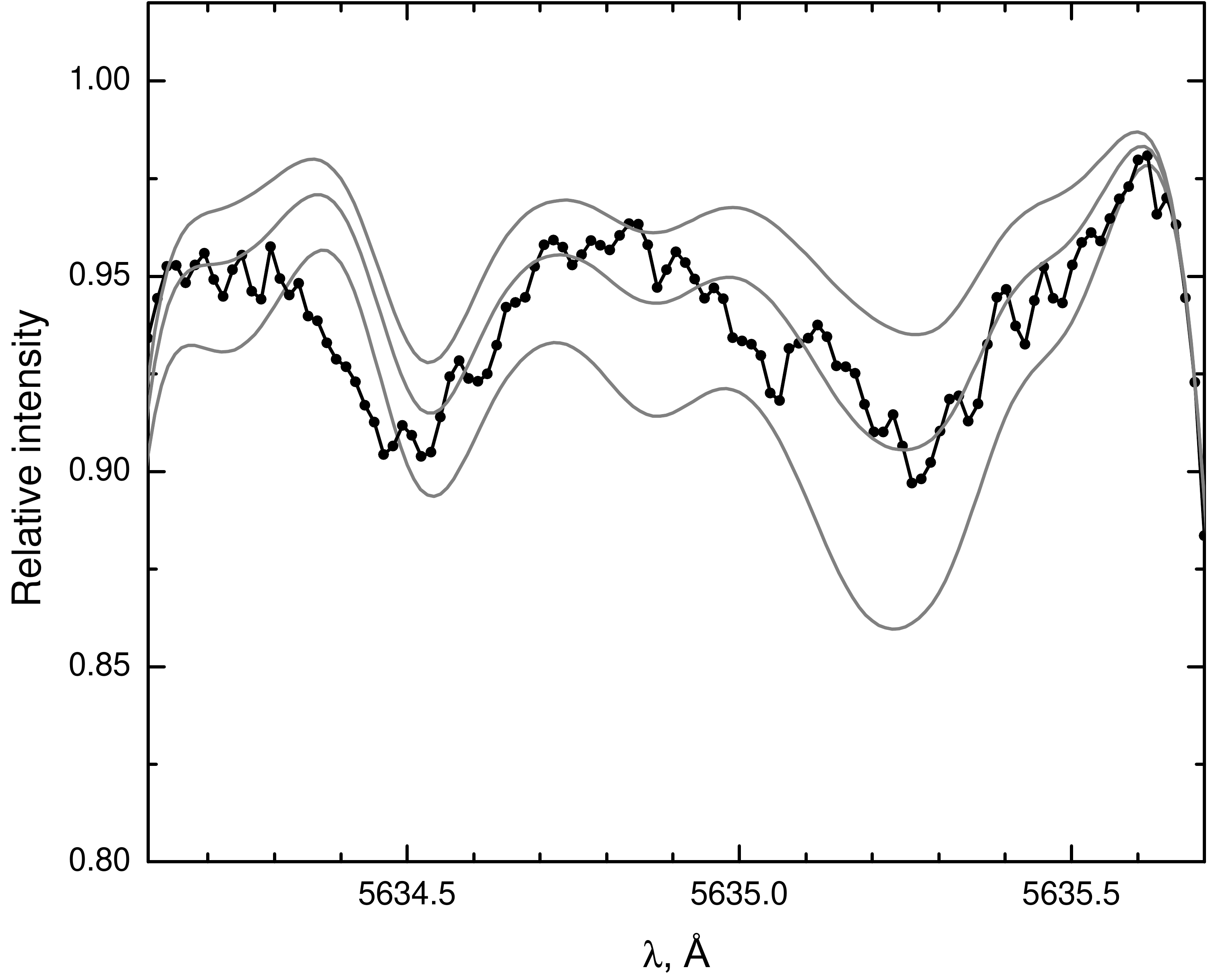}
\caption{ A fit to the ${\rm C}_2$ Swan (0,1) band head at 5635.5~{\AA} in the programme star NGC\,4815\,358. 
The observed spectrum is shown as a black line with dots. The synthetic spectra with ${\rm [C/Fe]}=-0.16 \pm 0.1$ 
are shown as grey lines.}
    \label{Fig8}
\end{figure}

\subsection{Evaluation of uncertainties}

The sensitivity of the abundance estimates $\Delta$[El/H] to changes in the atmospheric parameters are listed for the star NGC\,6705\,1625 in Table~4. 
Clearly, the abundances are not much affected when the parameter uncertainties quoted at the end of Sect.~2.2 are considered.

    \begin{table}
    \centering
       \caption{Effects on derived abundances, $\Delta$[El/H], resulting from model changes
for the star NGC\,6705\,1625. }
         \label{table:2}
       \[
          \begin{tabular}{lrrccc}
             \hline
             \noalign{\smallskip}
Species & ${ \Delta T_{\rm eff} }\atop{ \pm100~{\rm~K} }$ &
             ${ \Delta \log g }\atop{ \pm0.3 }$ &
             ${ \Delta v_{\rm t} }\atop{ \pm0.3~{\rm km~s}^{-1}}$ &
             ${ \Delta {\rm [Fe/H]} }\atop{ \pm0.1}$ &
             Total \\
             \noalign{\smallskip}
             \hline
             \noalign{\smallskip}
C\,(C$_2$) &    0.03 &    0.05 &    0.00 &    0.05 &    0.07    \\
N\,(CN) &    0.07 &    0.13 &    0.01 &    0.11 &    0.18    \\
O\,([O\,{\sc i}]) &    0.01 &    0.12 &    0.00 &    0.01 &    0.12    \\
                  \noalign{\smallskip}
             \hline
          \end{tabular}
       \]
    \end{table}

Since abundances of C, N, and O are bound together by the molecular equilibrium 
in the stellar atmosphere, we also investigated how an error in one of 
them typically affects the abundance determination of another. 
Thus $\Delta{\rm [O/H]}=0.10$ causes $\Delta{\rm [C/H]}=0.05$ and $\Delta{\rm [N/H]}=0.12$;   
$\Delta{\rm [C/H]}=0.10$ causes $\Delta{\rm [N/H]}=-0.17$ and $\Delta{\rm [O/H]}=0.02$; 
$\Delta {\rm [N/H]}=0.10$ has no effect on either the carbon or the oxygen abundances.

Random errors of abundance determinations, in this study mainly caused by uncertainties of the continuum placement and S/N ratios, can be evaluated 
from the scatter of abundances determined from different lines. The mean value of [C/H] scatter from the C$_2$ 5135 and 5635.5~{\AA} lines in Trumpler\,20 
is equal to $\pm 0.03$~dex, in NGC\,4815 $\pm 0.06$~dex, and in NGC\,6705 $\pm 0.02$~dex. The same low mean values of scatter are found in the case of [N/H] as well. 
Since oxygen abundances were determined just from one line at 6300.3~\AA, our ad hoc evaluation of random 
 errors for [O/H] is $\pm 0.05$~dex in NGC\,6705, and slightly larger, of about $\pm 0.15$~dex, in NGC\,4815. 

The influence of stellar rotation on the C, N, and O abundances in slowly rotating stars of our sample is very small; it becomes more important in stars rotating more rapidly. 
If $v\,{\rm sin}\,i$ is about 3 km\,s$^{-1}$, a difference of $\pm 1$~km\,s$^{-1}$ changes the abundances only by $\pm 0.01$~dex. 
If $v\,{\rm sin}\,i$ is about 7 km\,s$^{-1}$, a difference of $\pm 1$~km\,s$^{-1}$ changes the abundances of nitrogen and oxygen by about $\pm 0.06$~dex, 
while carbon abundances change only by $\pm 0.02$~dex. 
 
																
 \begin{table*}																	
     \centering																	
\begin{minipage}{150mm}																	
\caption{Main atmospheric parameters and elemental abundances for stars in the open cluster Trumpler\,20}																	
\label{table:1}																	
\begin{tabular}{ccccrrrc}																	
\hline\hline																	
\noalign{\smallskip}
GES ID	&	$T_{\rm eff}$	&	log~$g$	&	$v_{t}$	& [Fe/H] & [C/H] & [N/H] & C/N	\\
	&	K		&		&	km\,s$^{-1}$&		&		&						\\
\hline																			
\noalign{\smallskip}
12383595-6045245	&	5006	&	3.11	&	1.43	&	0.18	&	$-0.04$	&	0.46	&	1.26	\\
12383659-6045300	&	4863	&	2.84	&	1.23	&	0.21	&	0.04	&	0.60	&	1.10	\\
12385807-6030286	&	4488	&	2.13	&	1.48	&	0.00	&	$-0.20$	&	0.39	&	1.02	\\
12390411-6034001	&	4494	&	2.17	&	1.49	&	0.02	&	$-0.19$	&	0.43	&	0.95	\\
12390478-6041475	&	5017	&	2.99	&	1.39	&	0.13	&	$-0.11$	&	0.46	&	1.07	\\
12390710-6038057	&	5007	&	3.07	&	1.20	&	0.08	&	$-0.17$	&	0.53	&	0.79	\\
12391004-6038402	&	4516	&	2.17	&	1.57	&	$-0.04$	&	$-0.24$	&	0.39	&	0.93	\\
12391114-6036527	&	5003	&	3.09	&	1.34	&	0.20	&	$-0.02$	&	0.51	&	1.17	\\
12391201-6036322	&	4990	&	3.12	&	1.35	&	0.20	&	$-0.01$	&	0.54	&	1.12	\\
12391577-6034406	&	4873	&	2.99	&	1.37	&	$-0.03$	&	$-0.18$	&	0.35	&	1.17	\\
12392585-6038279	&	5086	&	3.24	&	1.39	&	0.14	&	$-0.08$	&	0.56	&	0.91	\\
12392637-6040217	&	4967	&	2.90	&	1.45	&	0.06	&	$-0.09$	&	0.54	&	0.93	\\
12392700-6036053	&	4830	&	2.70	&	1.46	&	0.06	&	$-0.11$	&	0.46	&	1.07	\\
12393132-6039422	&	5000	&	3.09	&	1.39	&	0.11	&	$-0.09$	&	0.60	&	0.81	\\
12393741-6032568	&	5030	&	3.10	&	1.38	&	0.14	&	$-0.09$	&	0.47	&	1.10	\\
12393782-6039051	&	4931	&	3.00	&	1.43	&	0.08	&	$-0.10$	&	0.56	&	0.87	\\
12394051-6041006	&	4976	&	2.94	&	1.42	&	0.15	&	$-0.09$	&	0.54	&	0.93	\\
12394122-6040040	&	5033	&	3.09	&	1.40	&	0.14	&	$-0.09$	&	0.53	&	0.95	\\
12394309-6039193	&	4889	&	2.92	&	1.39	&	0.13	&	$-0.13$	&	0.51	&	0.91	\\
12394387-6033166	&	5029	&	3.05	&	1.49	&	0.14	&	$-0.09$	&	0.63	&	0.76	\\
12394419-6034412	&	4990	&	3.03	&	1.40	&	0.17	&	$-0.13$	&	0.57	&	0.79	\\
12394475-6038339	&	4895	&	2.86	&	1.43	&	0.09	&	$-0.10$	&	0.58	&	0.83	\\
12394517-6038257	&	4955	&	3.00	&	1.36	&	0.22	&	$-0.02$	&	0.55	&	1.07	\\
12394596-6038389	&	4936	&	2.95	&	1.42	&	0.06	&	$-0.11$	&	0.44	&	1.12	\\
12394690-6033540	&	4984	&	2.97	&	1.42	&	0.08	&	$-0.10$	&	0.51	&	0.98	\\
12394715-6040584	&	4570	&	2.20	&	1.57	&	$-0.05$	&	$-0.27$	&	0.44	&	0.78	\\
12394742-6038411	&	4918	&	2.92	&	1.47	&	0.12	&	$-0.06$	&	0.57	&	0.93	\\
12394899-6033282	&	4919	&	2.78	&	1.49	&	0.05	&	$-0.12$	&	0.52	&	0.91	\\
12395426-6038369	&	4949	&	2.95	&	1.41	&	0.17	&	$-0.09$	&	0.54	&	0.93	\\
12395555-6037268	&	4923	&	2.93	&	1.34	&	0.11	&	$-0.09$	&	0.51	&	1.00	\\
12395569-6035233$^{*}$	&	5954	&	3.65	&	1.50	&	0.12	&	0.35	&		&		\\
12395655-6039011	&	4918	&	2.91	&	1.36	&	0.13	&	$-0.09$	&	0.44	&	1.17	\\
12395712-6039335	&	4964	&	3.00	&	1.40	&	0.09	&	$-0.08$	&	0.49	&	1.07	\\
12395975-6035072	&	4898	&	2.89	&	1.42	&	0.12	&	$-0.06$	&	0.57	&	0.93	\\
12400111-6031395	&	4910	&	2.99	&	1.37	&	0.13	&	$-0.10$	&	0.49	&	1.02	\\
12400116-6035218	&	4879	&	2.94	&	1.40	&	0.14	&	$-0.11$	&	0.41	&	1.20	\\
12400260-6039545	&	4503	&	2.12	&	1.53	&	$-0.03$	&	$-0.22$	&	0.36	&	1.05	\\
12400278-6041192	&	4993	&	3.13	&	1.33	&	0.19	&	$-0.02$	&	0.57	&	1.02	\\
12400451-6036566	&	4369	&	2.06	&	1.51	&	$-0.05$	&	$-0.27$	&	0.38	&	0.89	\\
12400755-6035445	&	4431	&	2.08	&	1.55	&	0.01	&	$-0.20$	&	0.43	&	0.93	\\
12402229-6037419	&	4969	&	2.95	&	1.46	&	0.09	&	$-0.09$	&	0.52	&	0.98	\\
12402480-6043101	&	4570	&	2.24	&	1.57	&	$-0.04$	&	$-0.25$	&	0.38	&	0.93	\\
\hline															
\noalign{\smallskip}															
Mean	&		&		&		&	0.10	&	$-0.10$	&	0.50	&	0.98	\\
s.d.	&		&		&		&	$\pm0.08$	&	$\pm0.10$	&	$\pm0.07$	&	$\pm0.12$	\\
\hline															
\end{tabular}												
\end{minipage}
\tablefoot{$^*$ This star is a subgiant located close to the main sequence turn-off and had no CN 
nands strong enough for the nitrogen abundance determination.}
\end{table*} 


																
 \begin{table*}																	
     \centering																	
\begin{minipage}{150mm}																	
\caption{Main atmospheric parameters and elemental abundances for stars in the open cluster NGC\,4815}																	
\label{table:1}																	
\begin{tabular}{ccccrrrrc}																	
\hline\hline																	
\noalign{\smallskip}
GES ID	&	$T_{\rm eff}$	&	log~$g$	&	$v_{t}$	& [Fe/H] & [C/H] & [N/H] & [O/H] & C/N	\\
	&	K		&		&	km\,s$^{-1}$&		&		&		&		&		\\
\hline																			
\noalign{\smallskip}
12572442-6455173	&	4328	&	1.88	&	1.67	&	0.01	&	$-0.14$	&	0.62	&	0.11	&	0.69	\\
12574328-6457386	&	4905	&	2.56	&	1.89	&	$-0.07$	&	$-0.19$	&	0.51	&	0.16	&	0.79	\\
12575511-6458483	&	4964	&	2.76	&	1.59	&	0.02	&	$-0.07$	&	0.57	&	0.22	&	0.91	\\
12575529-6456536	&	5062	&	2.68	&	1.13	&	$-0.04$	&	$-0.29$	&	0.43	&	$-0.02$	&	0.76	\\
12580262-6456492	&	4966	&	2.52	&	1.72	&	0.01	&	$-0.17$	&	0.52	&	0.13	&	0.81	\\
\hline																	
\noalign{\smallskip}																	
Mean	&		&		&		&	$-0.01$	&	$-0.17$	&	0.53	&	0.12	&	0.79\\	
s.d.	&		&		&		&	$\pm0.04$	&	$\pm0.08$	&	$\pm0.07$	&	$\pm0.09$	&	$\pm0.08$	\\

\hline
\end{tabular}												
\end{minipage}
\tablefoot{$^*$ This value was not included into the mean [O/Fe] value of the cluster stars.}
\end{table*}

 \begin{table*}																	
     \centering																	
\begin{minipage}{150mm}																	
\caption{Main atmospheric parameters and elemental abundances for stars in the open cluster NGC\,6705}																	
\label{table:1}																	
\begin{tabular}{ccccrrrrc}																	
\hline\hline																	
\noalign{\smallskip}
GES ID &	$T_{\rm eff}$	&	log~$g$	&	$v_{t}$	&	[Fe/H] & [C/H] & [N/H] & [O/H] & C/N	\\
	&	K		&		&	km\,s$^{-1}$&		&		&		&		&		\\
\hline																			
\noalign{\smallskip}
18502831-0615122	&	4543	&	1.82	&	1.83	& $-0.05$	&	$-0.14$	&	0.57	&	0.04	&	0.78	\\
18503724-0614364	&	4889	&	2.73	&	1.68	&	0.07	&	0.01	&	0.74	&	0.27	&	0.74	\\
18504563-0612038	&	4635	&	2.14	&	1.62	& $-0.07$	&	$-0.12$	&	0.56	&	0.13	&	0.83	\\
18504737-0617184	&	4330	&	1.71	&	1.64	& $-0.01$	&	$-0.10$	&	0.65	&	0.06	&	0.71	\\
18505494-0616182	&	4639	&	2.24	&	1.55	& $-0.05$	&	$-0.16$	&	0.54	&	0.08	&	0.79	\\
18505581-0618148	&	4496	&	1.94	&	1.73	&	0.07	&	0.04	&	0.54	&	0.16	&	1.26	\\
18505755-0613461	&	4828	&	2.55	&	1.40	&	0.02	&	$-0.14$	&	0.53	&	0.07	&	0.85	\\
18505944-0612435	&	4883	&	2.43	&	1.61	&	0.01	&	$-0.12$	&	0.65	&	0.11	&	0.68	\\
18510023-0616594	&	4351	&	1.85	&	1.75	& $-0.04$	&	$-0.03$	&	0.50	&	0.14	&	1.17	\\
18510032-0617183	&	4867	&	2.39	&	1.73	&	0.07	&	$-0.05$	&	0.66	&	0.15	&	0.78	\\
18510200-0617265	&	4321	&	1.80	&	1.60	& $-0.01$	&	$-0.01$	&	0.42	&	0.11	&	1.48	\\
18510289-0615301	&	4751	&	2.35	&	1.52	&	0.00	&	$-0.09$	&	0.55	&	0.12	&	0.91	\\
18510341-0616202	&	4798	&	2.16	&	1.89	&	0.06	&	$-0.05$	&	0.66	&	0.14	&	0.78	\\
18510358-0616112	&	4810	&	2.31	&	1.77	&	0.02	&	$-0.07$	&	0.59	&	0.18	&	0.87	\\
18510399-0620414	&	4660	&	2.09	&	1.71	& $-0.03$	&	$-0.12$	&	0.59	&	0.08	&	0.78	\\
18510662-0612442	&	4698	&	2.07	&	1.80	& $-0.02$	&	$-0.09$	&	0.55	&	0.12	&	0.91	\\
18510786-0617119	&	4767	&	2.48	&	1.87	&	0.06	&	0.03	&	0.70	&	0.25	&	0.85	\\
18510833-0616532	&	4713	&	2.19	&	1.83	&	0.05	&	$-0.04$	&	0.69	&	0.17	&	0.74	\\
18511013-0615486	&	4540	&	2.09	&	1.54	& $-0.07$	&	$-0.16$	&	0.65	&	0.14	&	0.62	\\
18511048-0615470	&	4688	&	2.06	&	1.81	&	$-0.14$	&	$-0.17$	&	0.59	&	0.06	&	0.69	\\
18511116-0614340	&	4332	&	1.74	&	1.67	&	0.01	&	$-0.08$	&	0.67	&	0.08	&	0.71	\\
18511452-0616551	&	4730	&	2.25	&	1.86	& $-0.04$	&	$-0.10$	&	0.53	&	0.16	&	0.93	\\
18511534-0618359	&	4675	&	2.18	&	1.85	&	0.08	&	$-0.05$	&	0.64	&	0.13	&	0.81	\\
18511571-0618146	&	4706	&	2.11	&	1.73	&	0.06	&	$-0.08$	&	0.64	&	0.10	&	0.76	\\
18512662-0614537	&	4422	&	1.97	&	1.62	&	0.03	&	$-0.06$	&	0.69	&	0.11	&	0.71	\\
18514034-0617128	&	4653	&	2.02	&	1.75	&	0.01	&	$-0.12$	&	0.65	&	0.13	&	0.68	\\
18514130-0620125	&	4687	&	2.13	&	1.73	&	0.01	&	$-0.10$	&	0.64	&	0.12	&	0.72	\\
\hline																	
\noalign{\smallskip}																	
Mean	&		&		&		&	$0.00$	&	$-0.08$	&	0.61	&	0.13	&	0.83\\	
s.d.	&		&		&		&	$\pm0.05$	&	$\pm0.06$	&	$\pm0.07$	&	$\pm0.05$	&	$\pm0.19$	\\
\hline
\end{tabular}												
\end{minipage}
\end{table*}

\section{Results and discussion}

The atmospheric parameters $T_{\rm eff}$, log\,$g$, $v_{t}$, [Fe/H] and abundances of C, N, and O chemical elements relative 
to iron [El/Fe] of the programme stars in Trumpler\,20, NGC\,4815, and NGC\,6705 are presented in Tables~5--7, respectively. 
They display the GESviDR2Final results. 
 
The abundances of oxygen were not determined for stars in Trumpler~20 because of blending by telluric lines. When determining abundances 
of carbon and nitrogen for stars of this cluster, we adopted a value of $-0.05$~dex for [O/Fe]. At the metallicity of this cluster, which is 
${\rm [Fe/H]}=0.10\pm0.08$ (s.d.) as determined from 42 Trumpler~20 stars in the GESviDR2Final (Table~5), this [O/Fe] value within about $\pm0.05$~dex uncertainty is typical 
(c.f. \citealt{Pagel95, Bensby14}).  A lower mean [O/Fe] value ($-0.18$~dex), based on an analysis of five clump stars,  
was presented for this cluster by \citet{Carraro14} even though the mean [Fe/H] value ($0.09\pm0.10$~dex) is very close to ours. 
We decided not to use this rather low [O/Fe] value from \citet{Carraro14} since the surface gravity log~$g$ values for the clump stars 
in their work are systematically lower than our values by about  0.3~dex. This difference in log~$g$ may cause the difference in oxygen abundances of about 0.12~dex (see Table~4). 
   
Preliminary analysis and interpretation of  $\alpha$- and iron-peak-element abundances in these three open clusters have 
been made by \citet{Magrini14}  on the basis of GESviDR1Final. It was concluded that the three clusters are essentially homogeneous in all investigated 
elements. As follows from our study, this conclusion is also valid for abundances of carbon, nitrogen, and oxygen.  
Tables~5--7 also contain mean values of [Fe/H] and [El/Fe], and it is seen that standard deviations (s.d.) do not 
exceed $\pm0.08$~dex. 

A sample of stars in Trumpler\,20 contains one subgiant located close to the main sequence turn-off (Fig.~1), 
i.e. Trumpler\,20\,430 (GES ID 12395569-6035233). Only the carbon abundance was determined for this star from quite 
 weak C$_2$ lines. All CN features were too weak for the nitrogen abundance determination. The abundance of carbon is 
by 0.45 dex larger in this unevolved star than in the investigated giants. It is expected that carbon abundances in unevolved stars are larger than in evolved, however, this large value of carbon in Trumpler\,20\,430 is probably not of primordial origin. 

The new mean GESviDR2Final [Fe/H] values for the investigated clusters agree with the GESviDR1Final results within the uncertainties.
An average iron abundance of 12 members of Trumpler~20  
${\rm [Fe/H]}=0.17\pm0.03$ (\citealt{Donati14}) is superceded by ${\rm [Fe/H]}=0.10\pm0.08$ based on 42 stars. 
In NGC\,4815, a mean [Fe/H] of NGC\,4815 was $+0.03 \pm 0.05$~ dex (\citealt{Friel14}). For the same 
five stars, the GESviDR2Final results yielded ${\rm [Fe/H]}=-0.01 \pm 0.04$~dex.  
An average iron abundance of 21 NGC\,6705 members in the GESviDR1Final was ${\rm [Fe/H]}=0.10\pm0.06$.  
A larger sample of 27  NGC\,6705 stars in GESviDR2Final gives ${\rm [Fe/H]}=0.0\pm0.05$. The new values were determined 
using advanced, in comparison to iDR1, atmospheric parameter homogenisation procedures (see \citealt{Smiljanic14} for details). 
The mean iron abundance of 10 K-giants in NGC\,6705, determined from high-resolution spectroscopy by \citet{Gonzalez00}, $0.07\pm0.05$~dex is close to our result. Eight of those stars are present in our sample.

\subsection{Carbon and nitrogen} 
 
Star clusters are recognised as the optimum test case for judging stellar evolutionary models and the validity of their 
physical assumptions. In our work, we will use the determined carbon and nitrogen abundance ratios  for this purpose. 
It is well known that canonical models, in which convection 
is the only driver of mixing inside a stellar interior, explain observations of stars in the lower part of the giant branch only.  Low- and intermediate-mass stars 
during the subsequent ascent on the RGB exhibit signatures of complex physical processes 
of extra-mixing that require challenging modelling.  In order to describe the observed surface abundances in different types of stars, various 
mechanisms of extra-mixing were proposed by a number of scientific groups
(see e.g. reviews by \citealt{Chaname05, Charbonnel06}, and papers by 
\citealt{Charbonnel10, Denissenkov10, Lagarde11, Palmerini11A, Wachlin11, Angelou12, Lagarde12, Karakas14} and references therein). 

\begin{figure}
\centering
\includegraphics[height = 8cm]{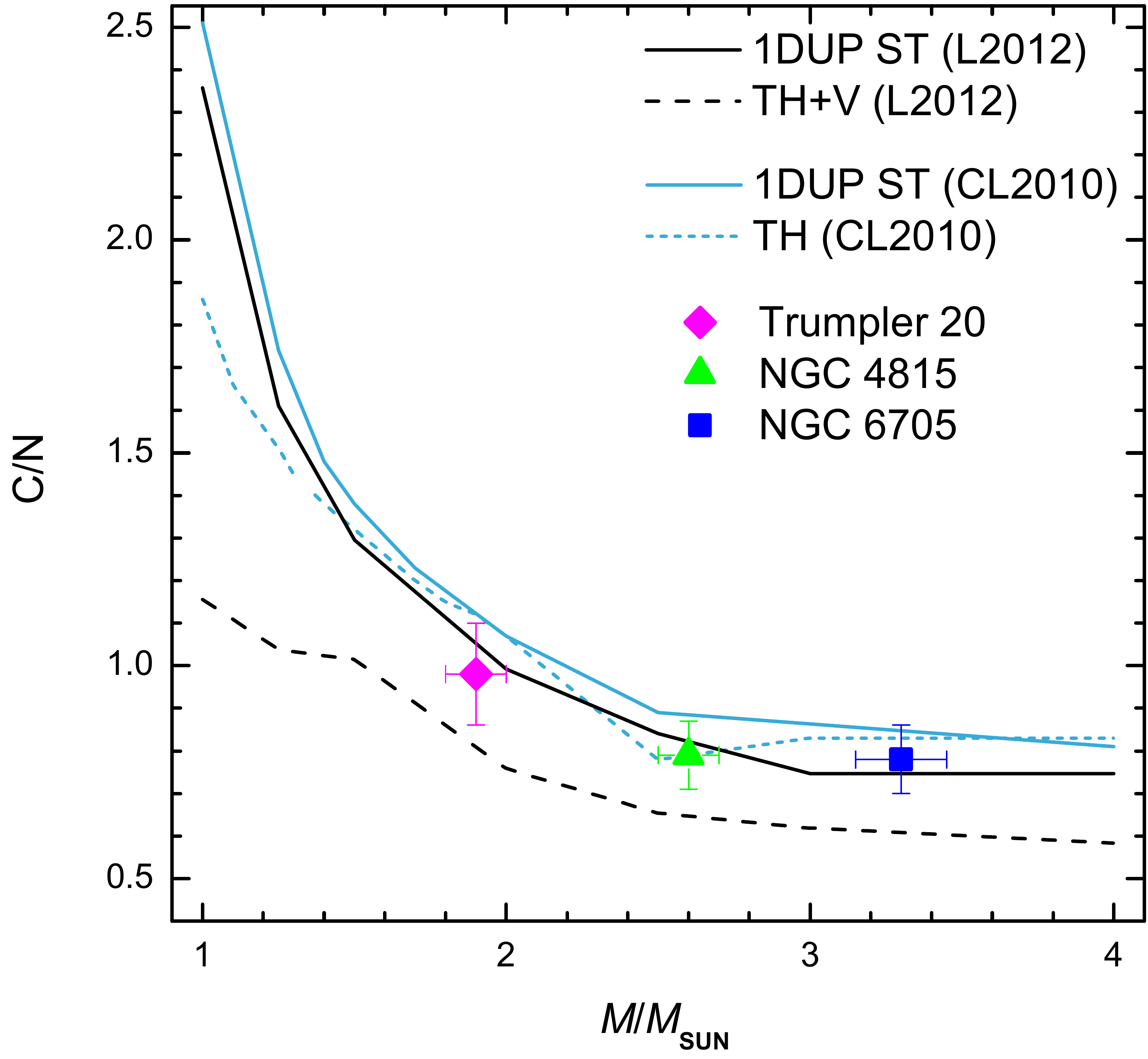}
\caption{Mean carbon-to-nitrogen ratios in stars of open clusters
as a function of stellar turn-off mass. 
The diamond represents Trumpler\,20 stars, the triangle is for NGC\,4815, and the square is for NGC\,6705 stars. 
The solid lines represent the C/N ratios predicted for stars at the first dredge-up with standard stellar evolutionary models by 
Charbonnel \& Lagarde (2010; blue upper line) and, more recently, Lagarde et al. (2012; black lower line). The blue dashed line shows the prediction 
when just thermohaline extra-mixing is introduced (Charbonnel \& Lagarde 2010), and the black dashed line is for the model that includes both 
the thermohaline and rotation induced mixing (Lagarde et al. 2012), see Subsect.~3.1 for more explanations. 
}
    \label{Fig9}
\end{figure}

The newest models with extra-mixing include a thermohaline instability induced mixing based on ideas of
\citet{Eggleton06} and \citet{Charbonnel07}.  \citet{Eggleton06} found
a mean molecular weight ($\mu$) inversion in their $1~M_{\odot}$ stellar evolution model, occurring after
the luminosity bump on the RGB, when the hydrogen burning shell reaches the chemically
homogeneous part of the envelope. The $\mu$-inversion is produced by the reaction
$^3{\rm He(}^3{\rm He,2p)}^4{\rm He}$, as predicted by \citet{Ulrich72}. It does not occur earlier, because the
magnitude of the $\mu$-inversion is small and negligible compared to a stabilizing
$\mu$-stratification. Following \citeauthor{Eggleton06}, \citet{Charbonnel07} computed stellar models
including the prescription by \citet{Ulrich72} and extended them to the case of a non-perfect gas for the turbulent
diffusivity produced by that instability in a stellar radiative zone. They found that a double diffusive instability
referred to as thermohaline convection, which had been discussed long ago in the literature (\citealt{Stern60}),
is important in the evolution of red giants. This mixing connects the convective envelope with the external wing of
the hydrogen burning shell and induces surface abundance modifications in red giant stars. 

Quantitative abundance values of mixing-sensitive chemical elements based on the thermohaline mixing model 
have been provided by \citet{Charbonnel10}.  In Fig.~9 we plotted a trend of the 
$1^{st}$ dredge-up C/N values computed using the standard (1DUP ST) model, as well as the trend of 
thermohaline induced extra-mixing (TH). 
The thermohaline mixing could be an important physical process governing the surface C/N ratios of stars with initial masses 
below 1.5~$M_{\odot}$, and its efficiency is increasing with decreasing initial stellar mass. The turn-off masses of the 
open clusters in our work are quite large (1.9, 2.6, and 3.3~$M_{\odot}$ in Trumpler\,20, NGC\,4815, and NGC\,6705, respectively).  
In comparison with the models of the $1^{st}$ dredge-up and 
the thermohaline mixing by \citet{Charbonnel10}, which are indistinguishable in this interval of stellar masses, 
the mean C/N ratios of the investigated clusters lie slightly lower, yet agree with them within the quoted uncertainties.   

A decrease of C/N values can also be caused by stellar rotation.   
\citet{Charbonnel10} computed models of rotation-induced mixing for stars at the zero-age main 
sequence (ZAMS) having rotational velocities of 110\,km\,s$^{-1}$, 250\,km\,s$^{-1}$, and 300\,km\,s$^{-1}$. 
The convective envelope was supposed to rotate as a solid body through the evolution. The transport
coefficients for chemicals associated with thermohaline- and rotation-induced mixing were simply added in the diffusion
equation and the possible interactions between the two mechanisms were not considered. The rotation-induced mixing
modifies the internal chemical structure of main sequence stars, although its signatures are revealed only later in
the stellar evolution when the first dredge-up occurs. More recently, \citet{Lagarde12} computed models with both the thermohaline and 
rotation induced mixing acting together. In Fig.~9 we show their model computed with standard (ST)
prescriptions, as well as the model including both thermohaline convection and rotation-induced mixing (TH+V). \citet{Lagarde12} also assumed 
solid-body rotation in the convective regions, however, in addition they assumed that the transport of angular momentum 
is dominated by the large amount of turbulence in these regions 
which instantaneously flattens out the angular velocity profile as it does for the abundance profiles. The initial rotation velocity 
of the models on the ZAMS was chosen at 45\% of the critical velocity at that point and  leads to mean velocities on the main 
sequence between 90 and 137~${\rm km\,s}^{-1}$. 
In Fig.~\ref{Fig9}, we can see that the C/N values in Trumpler\,20, NGC\,4815, and NGC\,6705 stars are not decreased 
as much as the model predicts if both the thermohaline and rotation induced extra mixing is at work. Indeed, the
observed C/N ratios are very close to predictions of the standard model at first dredge-up. 
In the already mentioned analysis of ten NGC\,6705 stars by \citet{Gonzalez00}, carbon isotope ratios were also determined. 
All the giants have $^{12}{\rm C}/^{13}{\rm C}\approx 20$, which also agrees with the $1^{st}$ dredge-up model. 
   
The thermohaline induced extra-mixing theory is under development. 
Magnetic activity also may play a role. \citet{Denissenkov09} investigated  a heat exchange between rising magnetic flux rings and their surrounding medium 
and proposed a model of magneto-thermohaline mixing.  
On the basis of three-dimensional numerical simulations of 
thermohaline convection, \citet{Denissenkov11} suggested that the salt-finger\footnote{The expression "salt-finger" comes from oceanology where 
thermohaline mixing is also widely used   
to model the regions of cooler, less salty water below the warmer water where the salinity is higher because of the evaporation 
from the surface. The so-called "fingers" of the warmer water penetrate the cooler water and the mixing occurs when the heat excess is 
exchanged (e.g., \citealt{Schmitt03, Ruddick03, Kunze03, Radko10}). }
spectrum might be shifted towards larger diameters by the toroidal magnetic field. \citet{Nucci14} investigated magnetic advection as a mechanism 
for deep mixing. According to their evaluation, in this case the mixing velocities are smaller than for convection, but larger than for diffusion and adequate 
for extra mixing in red giants. Unfortunately, these studies have not provided values of C/N that we could 
compare with observations.    
    
\citet{Wachlin11} computed full evolutionary sequences of RGB stars close to the luminosity bump and 
found that thermohaline mixing is not efficient enough for fingering convection to reach the bottom of the convective envelope 
of red giants. In order to reach the contact, the diffusion coefficient has to be artificially increased by about four orders of 
magnitude. 

A much larger, homogeneous data-base of CNO abundances in open clusters will be released in the framework of the Gaia-ESO Survey collaboration, which will significantly 
constrain mixing mechanisms in extant stellar evolutionary models. Unfortunately, carbon isotope ratios will not be 
investigated in this survey since there are no suitable spectral features in the selected spectral regions.

\subsection{Oxygen}

As described in the previous section, the abundances of C and N of the stars analysed in this work have been modified by stellar evolution processes and, hence, do not trace the initial composition of the stars anymore. The abundances of O, instead, still reflect the chemical composition of the stars at birth and can, thus, be used in studies of Galactic chemical evolution.

In Magrini et al. (2014), several abundance ratios ([Mg/Fe], [Si/Fe], [Ca/Fe], [Ti/Fe], [Cr/Fe], [Ni/Fe]) measured in the same clusters (GESviDR1Final) were compared with the predictions of two chemical 
evolution models (\citealt{Magrini09, Romano10}) and with field star abundance data. This comparison hinted at an inner birthplace for NGC\,6705. For NGC\,4815, the [Mg/Fe] ratio was also higher and similar to that in NGC\,6705.  
Here, we compare the [O/Fe] abundance ratios measured in NGC\,4815 and NGC\,6705 with the predictions of the same models (Fig.~10 and Fig.~11). 

In Fig.~10, we show the comparison with the [O/Fe] versus 
[Fe/H] trends predicted by Magrini et al. (2009) and by Romano et al. (2010) for different Galactocentric radii ($R_{\rm GC}$ equal to 4, 6, and 8~kpc). 
The data are consistent, within the errors,  
with the history of chemical enrichment of the solar neighbourhood and of the inner disc (R$_{\rm GC} \leq 4$~kpc). The trends predicted by Romano et al. (2010; Fig.~10, right panel) are 
almost independent of the Galactocentric radius, because of the adoption of an efficiency of star formation constant with the 
Galactocentric distance in their model. The trends by Magrini et al. (2009;  Fig.~10, left panel) vary with the Galactocentric distance as a result of an adopted  radial dependence 
of the star formation efficiency and of the infall rate. 

\begin{figure}
\centering
\includegraphics[height = 9cm]{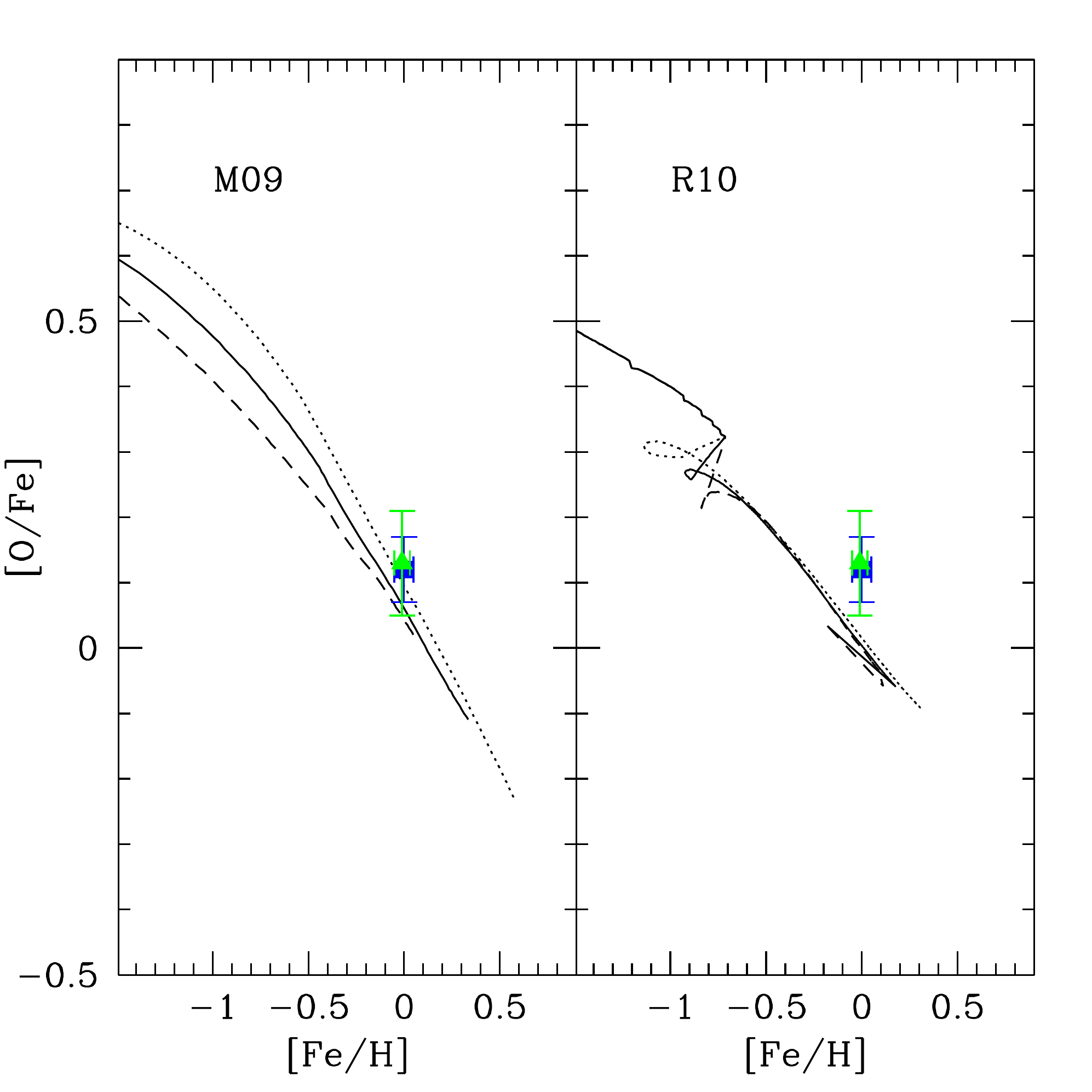}
\caption{[O/Fe] vs [Fe/H] in open clusters: NGC\,4815 (triangle) and NGC\,6705 (square). The curves are the theoretical predictions at three $R_{\rm GC}$: 4 kpc (dotted line), 6 kpc (continuous line), and 8 kpc (dashed line), for the model of Magrini et al. (2009, M09; left panel) and the model by Romano et al. (2010, R10; right panel). The theoretical ratios are normalised to the solar abundances predicted by each model. 
}
    \label{Fig10}
\end{figure}
\begin{figure}
\centering
\includegraphics[height = 9cm]{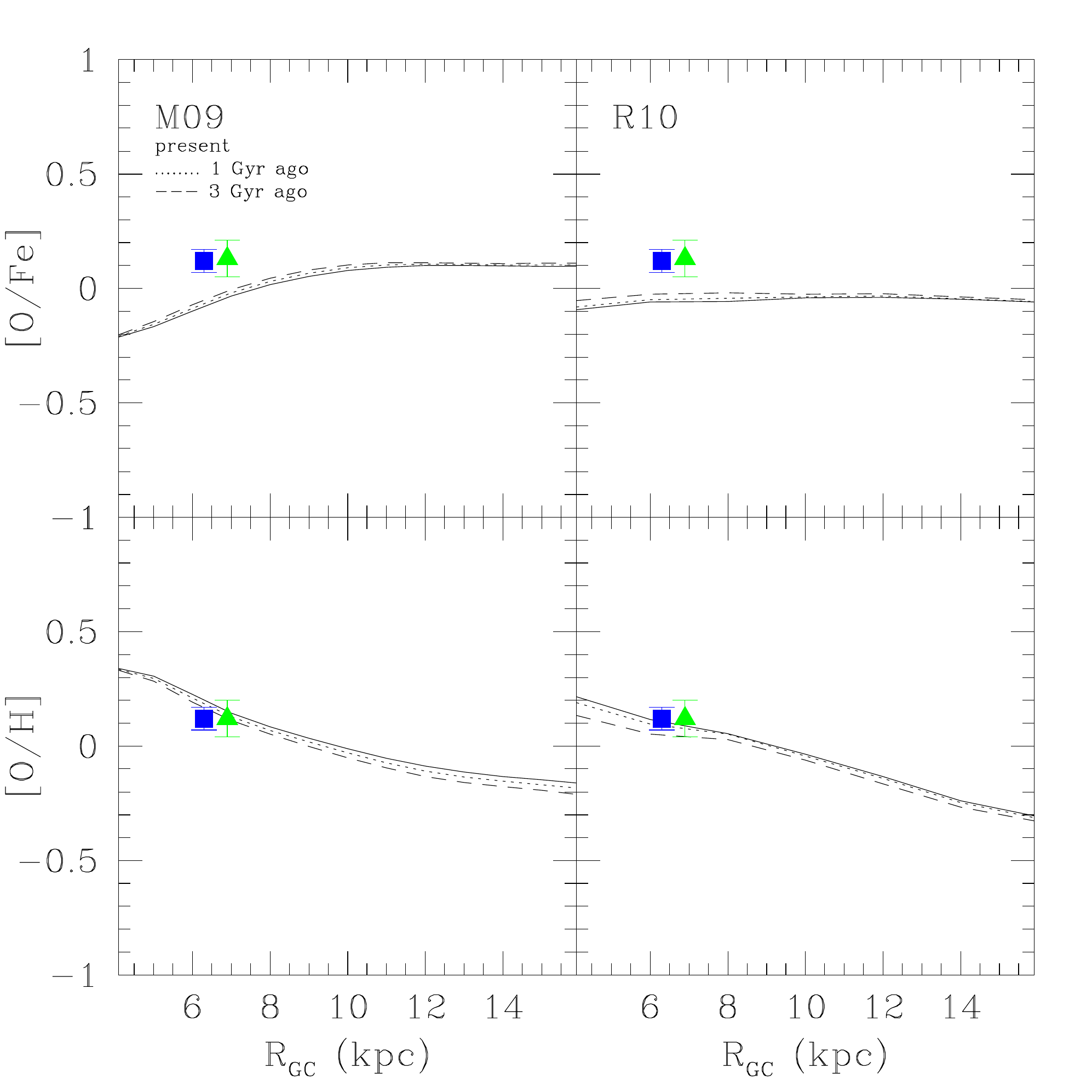}
\caption{[O/Fe]  and [O/H] vs R$_{\rm GC}$ in open clusters: NGC\,4815 (triangle) and NGC\,6705 (square). The curves are the theoretical predictions at three 
epochs: present time (continuous line), 1~Gyr ago (dotted line), and 3~Gyr ago (dashed line) for the model of Magrini et al. (2009, M09; left panel) and the model by Romano et al. (2010, R10; right panel). The theoretical ratios are normalised to the solar abundances predicted by each model. 
}
    \label{Fig11}
\end{figure}

The age of the clusters is hidden in this comparison, so    
in Fig.~11, we show the radial gradients of [O/Fe] and  [O/H] at three epochs: the present time, 1~Gyr ago, and 3~Gyr ago, i.e. in an age range 
comparable with the ages of the three clusters ($\sim$0.3, $\sim$0.5~Gyr and $\sim$1.5~Gyr for NGC\,6705, NGC\,4815, and Tr~20, respectively).   
Oxygen abundances are in good agreement with both models, while [O/Fe] is slightly higher than the model predictions. 

The comparison with the [O/Fe] versus 
[Fe/H] trends predicted by \citet{Magrini09} for different Galactocentric radii ($R_{\rm GC}$ equal to 4, 6, and 8~kpc; Fig.~10, left panel) seems to suggest an inner origin for both clusters. 
The curve for $R_{\rm GC}$= 4~kpc corresponds to a stronger infall rate and a higher star formation efficiency than assumed for $R_{\rm GC}$= 8~kpc. Thus, if one interprets the agreement between model predictions and observations as an indication of the conditions, in terms of star formation and infall rate, of the interstellar medium from which the clusters were born, he/she is led to conclude that their high [O/Fe] for their [Fe/H] indicate that they were born in a place subject to a more rapid enrichment than 
the solar neighbourhood, i.e. the inner disc. However, being more conservative, we notice that the data are not inconsistent, within the errors,  
with the history of chemical enrichment of the solar neighbourhood. The trends predicted by \citet{Romano10}; Fig.~10, right panel) are 
almost independent of the Galactocentric radius because of the adoption of an efficiency of star formation constant with the 
Galactocentric distance in their model. 

It is worth stressing at this point that the chemical evolution models adopted in this work do not take stellar migration into account. It is, however, unlikely that the predictions about the evolution of the heavy elements discussed in this paper are significantly modified by the process of stellar migration, since the parent stars would not travel very large distances before releasing their products, assuming typical velocities of 
1~km\,s$^{-1}$ ($\sim$1 kpc Gyr$^{-1}$; \citealt{Kordopatis13}) for the stars.

The oxygen abundance results obtained in this work for NGC\,4815 and NGC\,6705 will be utilised also for other aims of the Galactic evolution studies 
later on when a larger number of GES open clusters will be investigated. 

\section{Conclusions}

In this paper, we present the analysis of C, N, and O abundances for the first time in three open clusters observed in the Gaia-ESO Survey: Trumpler\,20, 
NGC\,4815, and NGC\,6705, in NGC\,4815. 
  
The C/N ratios in Trumpler\,20, NGC\,4815, and NGC\,6705 stars, which have turn-off masses of about 1.9, 2.6, and 3.3~$M_{\odot}$, 
are ${\rm C/N}=0.98\pm0.12$ (s.d.), ${\rm C/N}=0.79\pm0.08$, and ${\rm C/N}=0.83\pm0.19$, respectively.  The C/N values in the investigated 
clusters are not as low as predicted if thermohaline instability and rotation-induced mixing are at work in stars as suggested by \citet{Lagarde12}; 
rather, they are consistent with the predictions of the standard models from the same authors, or with models with only thermohaline-induced mixing 
as in \citet{Charbonnel10}.

We compare [O/H] and [O/Fe] abundance ratios with the results of two chemical evolution models: \citet{Magrini09}, and \citet{Romano10}.
The former includes a radial variation of the star formation efficiency and infall rate, while the latter assumes a constant star formation efficiency across the disc. 
The average values of  [O/Fe] vs. [Fe/H] in NGC\,6705 and NGC\,4815 are consistent, within the errors, with the clusters being born where they are found now in the disc. 
However, a comparison with the temporal evolution of the radial gradients of [O/Fe] and [O/H] show that they are slightly higher in [O/Fe] than expected 
for their age and present location. The [Mg/Fe] ratio in these clusters is also enhanced (\citealt{Magrini14}).

Many more open clusters will be investigated in the Gaia-ESO Survey, thus providing the homogeneous observational data needed to further develop both stellar 
and Galactic evolutionary models.

\begin{acknowledgements}
Based on data products from observations made with ESO Telescopes at the La Silla Paranal Observatory under programme ID 188.B-3002.
This work was partly supported by the European Union FP7 programme through ERC grant number 320360 and by the Leverhulme Trust through grant RPG-2012-541.
We acknowledge the support from INAF and Ministero dell' Istruzione, dell' Universit\`a' e della Ricerca (MIUR) in the form of the grant "Premiale VLT 2012" and  ``The Chemical and Dynamical Evolution of the Milky Way and Local Group
Galaxies'' (prot. 2010LY5N2T).
The results presented here benefit from discussions held during the Gaia-ESO workshops and conferences supported by the ESF (European Science Foundation) through the GREAT Research Network Programme.
T. Bensby was funded by grant No. 621-2009-3911 from the Swedish Research Council. T. Morel acknowledges financial support from Belspo for contract PRODEX GAIA-DPAC. R. Smiljanic was supported by the National Science Center of Poland through grant 2012/07/B/ST9/04428. 
This research has made use of the SIMBAD database, operated at CDS, Strasbourg, France, of NASA's Astrophysics Data System, 
of the compilation of atomic lines from the Vienna Atomic Line Database (VALD), and of the WEBDA database, operated at the 
Department of Theoretical Physics and Astrophysics of the Masaryk University.
\end{acknowledgements}

\end{document}